\documentclass[printer]{aa}

\usepackage{mathrsfs}
\usepackage{hyperref}

\usepackage{tikz}
\usepackage{subcaption}
\usepackage[]{algorithm2e}
\usepackage{txfonts}
\usepackage{nicefrac}
\usepackage{graphicx}

\renewcommand{\eqref}[1]{Eq.\,(\ref{#1})}
\usepackage{bbold}
\begin{document}

   \title{The Galactic 3D large-scale dust distribution via\\ Gaussian process regression on spherical coordinates}

   \author{R. H. Leike\inst{1}\inst{4}\inst{5}
          \and
          G. Edenhofer \inst{1}\inst{2}
          \and
          J. Knollm\"uller\inst{3}\inst{4}
          \and
          C. Alig\inst{2}\inst{4}
          \and
          P. Frank \inst{1}\inst{2}
          \and
          T. A. En\ss lin \inst{1}\inst{2}\inst{4}
          }

   \institute{Max Planck Institute for Astrophysics, Karl-Schwarzschildstra\ss e 1, 85748 Garching, Germany
         \and
         Ludwig-Maximilians-Universit\"at, Geschwister-Scholl Platz 1, 80539 Munich, Germany
         \and
         Technische Universität München, Boltzmannstr. 3, 85748 Garching, Germany
         \and
         Excellence Cluster ORIGINS, Boltzmannstrasse 2, D-85748 Garching, Germany
         \and
         This work was performed prior to employment by Amazon. Amazon does in no way endorse this publication.
       }

   \date{Received XXXX, accepted XXXX}


  \abstract{
  {Knowing the Galactic 3D dust distribution is relevant for understanding many processes in the interstellar medium and for correcting many astronomical observations for dust absorption and emission. }
   {Here, we aim for a 3D reconstruction of the Galactic dust distribution with an increase in the number of meaningful resolution elements by orders of magnitude with respect to previous reconstructions, while taking advantage of the dust's spatial correlations to inform the dust map.}
    {We use iterative grid refinement to define a log-normal process in spherical coordinates.
    This log-normal process assumes a fixed correlation structure, which was inferred in an earlier reconstruction of Galactic dust.
    Our map is informed through 111 Million data points, combining data of PANSTARRS, 2MASS, Gaia DR2 and ALLWISE.
    The log-normal process is discretized to 122 Billion degrees of freedom, a factor of 400 more than our previous map. We derive the most probable posterior map and an uncertainty estimate using natural gradient descent and the Fisher-Laplace approximation.}
   {The dust reconstruction covers a quarter of the volume of our Galaxy, with a maximum coordinate distance of $16\,\text{kpc}$, and meaningful information can be found up to at distances of $4\,$kpc, still improving upon our earlier map by a factor of 5 
   in maximal distance, of $900$ 
   in volume, and of about eighteen 
   in angular grid resolution.
   Unfortunately, the maximum posterior approach chosen to make the reconstruction computational affordable introduces artifacts and reduces the accuracy of our uncertainty estimate. }
   {Despite of the apparent limitations of the presented 3D dust map, a good part of the reconstructed structures are confirmed by independent maser observations.
   	Thus, the map is a step towards reliable 3D Galactic cartography and already can serve for a number of tasks, if used with care. }}


   \maketitle
%

\section{Introduction}
\label{sec:introduction}

Interstellar dust grains are conglomerations of particles ranging from the size of molecules to several micro-meters that are typically present in the cold interstellar medium.
They form from heavier elements which are more chemical active than hydrogen and helium, for example carbon and silicate \citep{draine2003interstellar}. %
As such, they only contribute a minor fraction of mass to the interstellar medium which is mostly composed of the less chemically active hydrogen and helium.
Dust plays an important role in the interstellar medium despite its small mass fraction.
It acts as a catalyst to facilitate the formation of molecular hydrogen from elemental hydrogen, a transition that is otherwise forbidden by conservation of angular momentum.
Dust also greatly enhances the cooling capacity of the interstellar medium due to it efficiently radiating at infrared wavelengths.
As such, it plays a role in star formation, and active star forming regions at the current epoch of the universe are associated with dense dust clouds.
At low temperatures, water and carbon dioxide ice can form on the surface \citep{burke2010ice}.
The heterogeneous composition of dust makes modelling its absorption properties challenging.
One way to model the wavelength dependent absorption of dust grains uses Mie theory,
which assumes that the scattering of photons is done by dielectric particles that are about the same size as the wavelength of photons, an approximation that works relatively well for dust grains.
A more modern approach to modelling the wavelength dependence of dust absorption is done by \citet{schlafly2016optical}, who use measurements to determine the wavelength dependence of the dust extinction and find a one parameter family that describes this extinction well.
The main observation of all models is that dust more efficiently scatters and absorbs higher wavelengths like blue light, while having a smaller cross-section for longer wavelengths.
Modelling dust absorption is important to understand how the presence of dust affects photon measurements here on Earth.
For example observations in the visible spectrum are heavily affected by dust extinction.
Three dimensional reconstructions of dust extinctions play a special role.
On the one hand, they allow to disambiguate observations.
For example two astrophysical components that overlap on the sky might actually be separated by their distance, and estimating the distance to dust clouds can help to establish or solidify connections that cannot be made from angular maps alone \citep{leike2021optical,bialy2021per}.
Constraining distances to dust clouds is also of interest to understand star forming regions \citep{zucker2020compendium}, as it allows to better understand the environment around the young starts.

There are several recent approaches to three dimensional extinction mapping.
Several of them choose Cartesian coordinates \cite{lallement2019gaia, lallement20183d, capitanio2017three, leike2020resolving,leike2019charting}.
These do not favor certain points or directions in space, making it easier to incorporate isotropic and homogeneous correlations into the prior statistics.
For Cartesian grids, the angular resolution can suffer, as the voxels of nearby dust structures appear on large angular scales in sky projection.
This is problematic, as nearby dust clouds obscure any line of sight (LOS) passing through them, while the angular coarse grid representing of them is unable to display their small-scale angular features.
Those, however, are often well probed by absorption measurements from the large number of distant stars.
As a consequence of this, Cartesian grid reconstructions can have artifacts throughout their entire volume.
This is because of their inability to represent nearby small-scale features requested by the data, which then might be represented instead by spurious structures at larger distances, where the voxel size corresponds to smaller angular scales.

There are other approaches that circumvent this angular resolution problem of Cartesian grids by placing their resolution elements on every LOS in a directions of an observed star \citep{kh2018detection, kh2017inferring, kh2020detailed}.
This ensures that the grid resolution is sufficiently high to capture the data resolution in angular directions.
However, these approaches inflate the costs of computing correlations between all points.
Recent advances \citep{kh2020detailed} have overcome this limitation to some degree by iteratively reconstructing shells of dust with increasing distance.

A spherical coordinate system centered on the Earth can circumvent the problem of changing angular diameters of Cartesian voxels.
Spherical coordinates are already in frequent use for three dimensional dust mapping \citep{green2018galactic, green20193d, chen2018three}.
For example, in \citet{green20193d} the sky is decomposed into small angular cells with exponentially increasingly radial sizes.
On this grid, a one dimensional Bayesian analysis is performed along the individual LOS.
\citet{chen2018three} use linear distance bins and random forest regression to obtain a 3D dust extinction map. 
Furthermore, in \citet{green20193d}, spherical grids are used, but correlations are taken into account only along a LOS in the first stages of the algorithm.
Later stages of the algorithm then use information from neighbouring lines of sight and assume a fixed correlation kernel that falls exponentially after $1.5\,\text{pc}$, in an approach that resembles Gibbs sampling.

The approach presented in this work is similar to those discussed above by also using a spherical coordinate system, but it takes into account correlations perpendicular to the LOS as well as along the LOS over larger distances.
Such long range correlations in the dust distribution were observed to be present in the 3D dust reconstructions by \cite{leike2020resolving}, and exploiting them in a reconstruction is helping significantly to reduce the distance uncertainty for any tomographically probed structure.
However, it turns out that due to the here for computational cost reasons used maximum a posteriori (MAP) approach, our uncertainty estimates, i.e. how well the actual dust density is captured by the uncertainty, is much less precise than that of reconstructions taking a sampling based approach, i.e. as used by \citet{green20193d}.

We split the sky into 424 patches of about $12.5^\circ\times12.5^\circ$ accounting for correlations within one patch, and propagate information on the boundaries to neighbouring patches between iterations of our algorithm.
Our correlations are introduced through a Gaussian process kernel based on the power spectrum of the logarithmic dust extinction from \cite{leike2020resolving}.
In this work, we use iterative grid refinement \citep{edenhofer2021sparse} to apply the two point correlation kernel on the spherical coordinates of the angular patches, giving rise to a generative model for correlated dust within one patch.
For details about how the grid refinement within grid cones associated with the angular patches is done and how
the boundary conditions between the different angular cones are implemented we refer to \autoref{sec:grid-refinement-for-cones} and \autoref{sec:boundary-conditions}, respectively.

\section{Data}
\label{sec:data}

We use data from \citet{anders2019photo}, who combine Gaia DR2, ALLWISE, PANSTARRS, and 2MASS measurements. %
They use the \verb|StarHorse| \cite{queiroz2018starhorse} pipeline, which constrains stellar parameters through photometric data of multiple energy bands in a Bayesian fashion.
We employ the same data quality selection criteria as in our previous reconstruction \citep{leike2020resolving}, i.e.
\begin{align}
    \text{SH\_OUTFLAG} &= \text{00000},\\
    \text{SH\_GAIAFLAG} &= \text{000},\\
    \text{ph} &\in \text{Table } \ref{table:good-photoflags} ,\\
    \nicefrac{\sigma_\omega}{m_\omega} &< 0.3,\text{ and}\\
    \text{av}_{05} &\neq \text{av}_{16} \,.
\end{align}
Furthermore, we select stars that are inside the reconstructed volume with $84\%$ credibility, i.e. stars for which
\begin{align}
	\text{dist}_{16}&>40\,\text{pc and} \\
	\text{dist}_{84}&<16\,000\,\text{pc.}
\end{align}
These criteria select $110\,983\,305$ stars, about a factor of $22$ more data than was used for \cite{leike2020resolving}.

\section{Algorithm}

Our algorithm is inspired by Bayesian inference.
The central quantity of Bayesian inference is the posterior probability density
$P(\rho|\mathcal{D})$, which in our case is the probability of a dust extinction density field $\rho$ given the data $\mathcal{D}$.
By Bayes theorem, we can decompose this probability density as
\begin{align}
    P(\rho|\mathcal{D}) = \frac{P(\mathcal{D}|\rho) P(\rho)}{P(\mathcal{D})}\ ,
\end{align}
where $P(\mathcal{D}|\rho)$ is the likelihood of the data $\mathcal{D}$ given the density $\rho$ and $P(\rho)$ is the prior on the density $\rho$.
The normalization $P(\mathcal{D})$ is often hard to calculate, but as we use a MAP approach, not necessary here.

To express the prior $P(\rho)$ we employ a generative model, i.e. instead of defining $P(\rho)$ directly we use latent parameters $\xi$ that are connected to $\rho$ through a generative process $\rho(\xi)$.
All the complexity of the prior is absorbed in this functional relationship $\rho(\xi)$, and we can assume $\xi$ to be an independent normal distributed random vector \citep{knollmuller2018encoding}.
This aids with convergence and can also influence the MAP, since MAP is not a coordinate invariant method.

Our data can be split into independently measured sub-data sets $\mathcal{D}_i$ with individual likelihoods, i.e. $\mathcal{D}=(\mathcal{D}_i)_i$.
Therefore, the overall likelihood is given by a product of all these individual likelihoods,
i.e.
\begin{align}
    P(\mathcal{D}|\rho) = \prod_{i=1}^n P(\mathcal{D}_i|\rho)\ .
\end{align}
We can decompose this probability into a product because the noise statistics of individual measurements are independent.
The individual data sub-sets $\mathcal{D}_i=(a_i,\omega_i)$ consist of a parallax estimate $\omega_i$ and an extinction estimate $a_i$ for each individual star $i$.
We denote the true extinction of star $i$ by $A_i$, which will differ from its measurement based estimate $a_i$ due to measurement noise and ambiguities in the extinction estimate.
We further split any star-likelihood $P(a_i,\omega_i|\rho)$ into two parts, a mean response $R_i(\rho)$, which states what the expected extinction for star $i$ is given $\rho$, but not knowing its distance $d_i$ precisely, and a noise statistic $P(a_i|R_i(\rho))$, which quantifies how likely the measured extinction $a_i$ is given the expected one.
These quantities are constructed in the following subsections.

\subsection{Response}
\label{sec:response}

In this subsection we discuss how to compute the expected extinction $R_i(\rho)$ given a three dimensional dust density $\rho$ and an estimate for the star distance $d_i$.

Our data contains measurements of the integrated extinction $A$ of stars.
The integrated extinction $A(r,\phi)$ at any location in solar centered spherical coordinates with distance $r$ and normalized direction $\phi$ is related to the differential extinction density $\rho$ via
\begin{align}
    A(r,\phi) = \int_0^{r}\text{d}r^\prime\rho(r^\prime\phi) \ . \label{eq:cumsum}
\end{align}

If we would know the distance $d_i$ to a star $i$ in direction $\phi_i$, we would expect its measured extinction $a_i$ to be equal to $A(d_i,\phi_i)$ up to noise.
In practice the distance $d_i$ is not known exactly.
We calculate the expected extinction $R$ given the measured parallax $\omega_i$ of the star:
\begin{align}
    R_i(\rho) &= \mathbb{E}_{P(d_i|\omega_i)}\left[A(d_i,\phi_i)\right]\nonumber\\
    &= \mathbb{E}_{P(d_i|\omega_i)}\left[\int_0^{d_i}\text{d}r\rho(r\phi_i)\right] \label{eq:los-expectation}
\end{align}
We discretize the dust extinction distribution using spherical coordinates, thus $A$ can be calculated from $\rho$ using a weighted cumulative sum along the radial grid axis.
The expectation value in \autoref{eq:los-expectation} for discretized formulation then corresponds to a dot product between a binned distance probability distribution $P(r_j < d_i \leq r_{j+1}|\omega)$ and the corresponding differential dust opacities for these distances $ \rho(\phi)_j$: %
\begin{align}
    \mathbb{E}_{P(d_i|\omega_i)}\left[\int_0^{d_i}\text{d}r\rho(r\phi_i)\right] \approx \sum_j P(r_j < d_i \leq r_{j+1}|\omega)\, \rho(\phi)_j \label{eq:los-expectation-discretized} .
\end{align}
Here,  $r_j$ and $r_{j+1}$ are the boundaries of the $j$-th distance bin and $\rho(\phi)_j$ is the discretized dust density for direction $\phi$ and at the $j$-th distance bin.
Both operations described by \autoref{eq:cumsum} and \autoref{eq:los-expectation-discretized} can be computed efficiently, allowing to take vastly more data into account compared to what was possible for our previous maps \citep{leike2020resolving}.
There, the non-alignment of lines of sight with our Cartesian grid axes required accessing computer memory in an inefficient way.

\subsection{Noise statistics}
\label{sec:noise-statistics}

We have discussed in Sec.\,\ref{sec:response} how to calculate the expected amount of dust extinction to a star $i$.
In this section we model how close this estimate is to the measured G-band extinction $a_i$ for that star, given we would know the true three dimensional distribution of dust extinction.
There are two different sources of noise that we consider.
On the one hand, the measurement of extinction is intrinsically uncertain, and we model this by assuming the true extinction to be Student-T distributed given the measured extinction.
The Student-T distribution is robust to outliers, which have been found in our dataset.

On the other hand, we incur an error because we only calculate the expected extinction with respect to the unknown position of the star. This uncertainty about the stars position introduces an uncertainty about its extinction even if the true dust distribution is given.

\begin{figure}[ht]
    \includegraphics[trim={0.0cm .0cm .2cm .0cm}, clip, width=.49\textwidth]{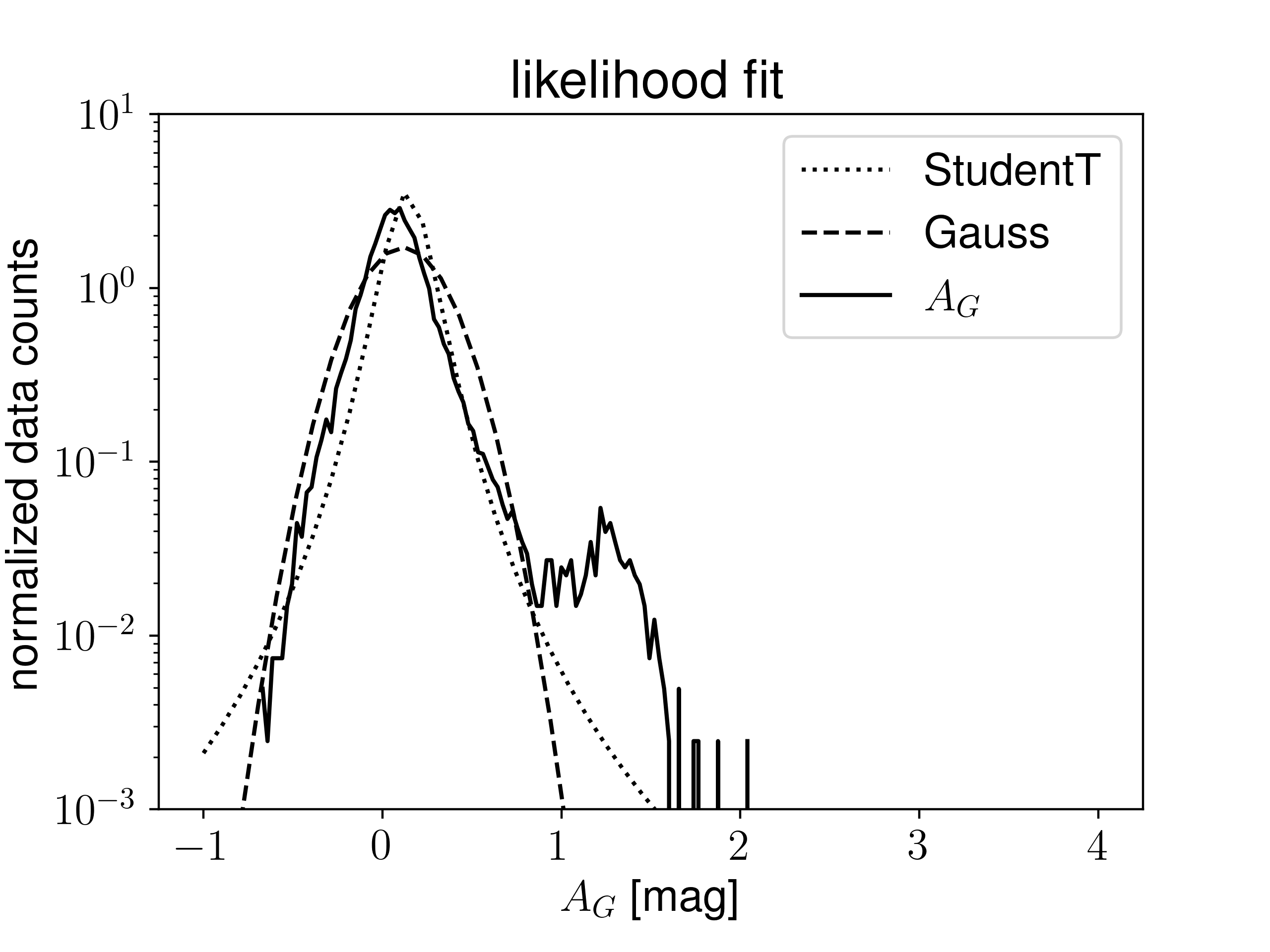}
    \caption{\label{fig:fitting-likelihood}
    The solid line shows data counts for the "GBPRPgrizyJHKsW1W2" photoflag for lines of sight in dustless regions.
    We fitted this with a Student-T distribution shown dotted, using a maximum likelihood estimation.
    The dashed line is a fit using a Gaussian distribution for comparison.
    The x-axis shows logarithmic density, the y-axis is G-band extinction in magnitudes.
        }                       %
\end{figure}

To quantify the first effect we look at extinction estimates for stars in dustless regions.
We fit the probability of finding a particular extinction estimate given the true extinction of $0$ with a Student-T distribution.
We obtain the parameters for the Student-T distribution through a maximum likelihood fit, using separate parameters for each so called photoflag of the data values.
This is in contrast to \cite{leike2020resolving}, where we assumed a Gaussian distribution of the estimated extinction given the true extinction.
The Student-T distribution is more robust to outliers and seems to be a better fit to the data overall.
We show how a Student-T and a Gaussian distribution fit the photoflag with the highest amount of data points in \autoref{fig:fitting-likelihood}

The photoflag determines which energy bands of which instruments were used to estimate the extinction of the corresponding star.
We expect the uncertainty of the extinction estimate to depend on the amount of energy band measured: The more energy bands were measured, the less uncertainty is expected.
This trend can indeed be seen in Table\,\ref{table:good-photoflags}, where we show the obtained parameters of the Student-T distributions for each of the photoflags.
With the fitted parameters $m_\text{ph}$, $\sigma_\text{ph}$, and $\nu_\text{ph}$, our likelihood of the $i$-th measured extinction given the true dust distribution $\rho$, the photoflag $\text{ph}$ of the datapoint, the photoflag $\text{ph}$ of the datapoint, and the true distance $d_i$ and direction $\phi_i$ is
\begin{align}
    P(a_i|A(d_i,\phi_i)) = T(a_i|A(d_i,\phi_i) + m_{\text{ph}(i)}, \sigma_{\text{ph}(i)}, \nu_{\text{ph}(i)})\ ,\label{eq:noise-statistic}
\end{align}
where $T$ is the Student-T distribution that is defined through its logarithmic probability density function
\begin{align}
    -\text{ln}\left[T(x|m, \sigma, \nu)\right] =
    \frac{\nu+1}{2}\text{ln}\left(1+\frac{x^2}{\sigma^2\nu}\right) + \text{ln}(\sigma) + \iota(\nu) \ ,
\end{align}
where $\iota(\nu)$ is a constant in $x$ that is irrelevant for our inference.
\eqref{eq:noise-statistic} uses the true distance $d_i$ instead of the expected distance given the measured parallax $\omega_i$.
The true distance $d_i$ is unknown, and when using the expected extinction given the measured parallax $\omega_i$, we incur an additional error, as discussed before in \autoref{sec:noise-statistics}.
To get an estimate of how large this effect is, we calculate the variance of the extinction with respect to a given dust distribution but unknown star position, i.e. the expected squared deviation of integrated extinction $A(d_i, \phi)$ and expected extinction  $R_i(\rho)$.
This can be expressed as
\begin{align}
    R^\text{var}_i(\rho)
     &= \mathbb{E}_{P(d_i|\omega_i)}A^2(d_i,\phi_i) - \mathbb{E}_{P(d_i|\omega_i)}\left[A(d_i,\phi_i)\right]^2. \label{eq:noise-correction}
\end{align}
In our previous reconstruction \citep{leike2020resolving} we calculated this supplementary variance $R^\text{var}$ by sampling different distances $d_i$ and taking the sample standard deviation as noise correction.
Here, we simply calculate its effect directly, using \eqref{eq:noise-correction}, as this leads to a more stable estimate.

We add the supplementary variance to the squared scale parameter of the Student-T distribution, such that our overall likelihood for the $i$-th data point becomes
\begin{align}
    P(a_i|R^*_i(\rho)) = T\left(a_i\left|R^\text{avg}_i(\rho) + m_{\text{ph}(i)}, \sqrt{\sigma^2_{\text{ph}(i)}+R^\text{var}_i(\rho)}, \nu_{\text{ph}(i)}\right.\right)\ .
\end{align}
This correction should eliminate systematic errors from the parallax uncertainty completely in a fully Bayesian analysis, at the cost of neglecting some of the information contained in the data.
The implicit assumption that enters the correction is that the star distance estimates are independent from the stars' extinctions.
However, given that the three dimensional dust distribution couples the estimates of quantities in practice some correlation between their measurement errors is to be expected.
This correlation is neglected here.
In the future, more sophisticated algorithms hopefully will take this correlation between the measured quantities into account and therefore will yield better results than we are able to achieve here.

\begin{table}
\centering
\footnotesize
\begin{tabular}{cccc}
    $\text{ph}=\,$SH\_PHOTOFLAG & $m_\text{ph}$ & $\sigma_\text{ph}$ & $\nu$\\
\hline
GBPRP & 0.415 & 0.320 & 3.687\\
GBPRPJHKs & 0.104 & 0.172 & 3.399\\
GBPRPJHKs\#W1W2 & 0.315 & 0.538 & 314115.072\\
GBPRPJHKsW1W2 & 0.089 & 0.125 & 2.537\\
GBPRPgrizyJHKs & 0.173 & 0.105 & 2.279\\
GBPRPgrizyJHKsW1W2 & 0.140 & 0.102 & 2.951\\
GBPRPiJHKsW1W2 & 0.078 & 0.131 & 3.072\\
GBPRPiyJHKsW1W2 & 0.131 & 0.110 & 2.522
\end{tabular}
\normalsize
    \caption{SH\_PHOTOFLAG values and the corresponding mean $m_\text{ph}$, scale parameter $\sigma_\text{ph}$, and degree of freedom parameter $\nu$ for stars in directions of voids of dust. We classify regions as void of dust if and only if the Planck dust map shows weaker emission than $\text{exp}(2)\nicefrac{\mu K}{rJ}$, analogously to \cite{leike2020resolving}.}
\label{table:good-photoflags}
\end{table}

\section{Prior}
\label{sec:prior}

\subsection{Mathematical considerations}

The most important a priori information that we exploit is that the dust density is positive, correlated, and varies over multiple orders of magnitude.
All these properties can be realized by modelling the logarithmic dust extinction density as a Gaussian Process (GP),
\begin{align}
	\rho(\xi) =\ &\rho_0 \exp(\tau(\xi)) \,,
	\\ \tau \curvearrowleft\ &\mathcal{G}(\tau|0,C)
\end{align}
with $\rho_0=\nicefrac{1}{1000}\nicefrac{\text{mag}}{\text{pc}}$ denoting the prior median extinction density and $C$ the correlation kernel of the logarithmic dust extinction $\tau$.

To encode that no point and/or direction is a priori special, we use an isotropic and homogeneous kernel for our GP.
Thus the a-priori assumed correlation $C(x,x^\prime)$ of the logarithmic dust density at two different points $x$ and $x^\prime$ depends only on the distance $r=\|x-x^\prime\|$ of these points.
This assumption is equivalent to requiring the kernel to arise from an isotropic power spectrum $\text{P}_k$ according to the Wiener-Khinchin theorem \citep{Wiener1930generalized}:
\begin{align}
    C(x) = \int_{-\infty}^\infty \text{d}^3k\, \text{exp}(-ikx)\text{P}_k \label{eq:kernel-from-power}\ ,
\end{align}
with isotropy implying that
\begin{align}
    \forall x,x^\prime :\|x\| = \|x^\prime\| &\Rightarrow C(x) = C(x^\prime)\\
    \forall k,k^\prime :\|k\| = \|k^\prime\| &\Rightarrow \text{P}_k = \text{P}_{k^\prime}\ .
\end{align}

\subsection{Practical implementation}

We use a fixed kernel and iterative grid refinement to construct the generative model that encodes our prior knowledge.
Thus, in contrast to our previous dust reconstruction \citep{leike2020resolving}, here we do not reconstruct the power spectrum.
Instead we use the reconstruction of \citet{leike2020resolving} and fit a truncated power law to the spectrum found previously.
This enables us to inter-and extrapolate the power spectrum to all scales appearing in our reconstructions.

We parametrize the power spectrum as a power law at large Fourier-scales with a free zero mode,
\begin{align}
    \text{P}_k = z_0 \delta(k) + \frac{a}{1+\left(\|k\|/k_0\right)^\alpha}\ , \label{eq:power-spectrum-form}
\end{align}
with parameters $z_0$, $a$, $k_0$, and $\alpha$.
Details on the fitting procedure can be found in \autoref{sec:fitting-the-kernel}.

With this power spectrum, we employ iterative grid refinement \citep{edenhofer2021sparse} to construct a generative model of the logarithmic dust extinction realization $\tau$ inside an angular patch with dimension of $12.5^\circ\times12.5^\circ$ given the latent standard normal distributed variable $\xi$ that determines $\varrho = \tau(\xi)$.
I.e. iterative grid refinement provides us with an algorithm running in $\mathcal{O}(\text{dim}(\tau))$ to calculate $\tau(\xi)$ such that
\begin{align}
	\tau \curvearrowleft\ &\mathcal{G}(\tau|0,C)\ ,
\end{align}
approximately by sampling on a coarser grid and then iteratively increasing the resolution to reach the target resolution.
More details on the procedure are explained in \autoref{sec:grid-refinement-for-cones}.
We cover the whole sky using many patches, letting individual patches overlap for about $2.5^\circ$.
To propagate information between the patches we add an additional loss penalty that punishes differences in the logarithmic dust realizations at the boundaries of each patch, as described in \autoref{sec:boundary-conditions}.

\section{Inference}
\label{sec:inference}

We maximize the posterior probability density $P(\xi|\mathcal{D})$ with respect to the latent variable $\xi$.
This is achieved by using a second order minimization scheme that minimizes the loss function $\mathscr{L}$.
Up to irrelevant constants this loss function is equal to the negative logarithmic posterior probability:
\begin{align}
    \mathscr{L}(\xi) = \sum_i-\text{ln}(P(\mathcal{D}_i|\xi)) + \sum_j\frac{1}{2}\xi_j^2\ , \label{eq:overall-loss}
\end{align}
where $\sum_j\frac{1}{2}\xi_j^2$ is the negative log-prior, which takes this simple quadratic form (that specifies $\xi \curvearrowleft \mathcal{G}(\xi|0,\mathbb{1})$) because we absorbed all the complexity into the generative model $\rho(\xi)$.

Our second order Newton-like optimization scheme uses the inverse of the Bayesian Fisher metric
\begin{align}
    F(\xi) = \mathbb{E}_{P(\mathcal{D}|\xi)}\left[ \frac{\partial \text{ln}(P(\xi|\mathcal{D}_i))}{\partial \xi}\frac{\partial \text{ln}(P(\xi|\mathcal{D}_i))}{\partial \xi^T}\right]\ ,
\end{align}
as a preconditioner to the gradient of our loss function, in order to determine the direction for the next step.
Thus, a single minimization step takes the form
\begin{align}
    \xi_i \leftarrow \xi_{i-1} - \alpha F^{-1}\frac{\partial \mathscr{L}}{\partial \xi}\ ,
\end{align}
where $\alpha$ is the step length, which is determined by a line search algorithm. The inverse of $F$ is applied to the gradient by the Conjugate Gradient algorithm \citep{hestenes1952methods}.
Note that the application of the Fisher metric to a vector can be computed directly using auto-differentiation on the generative model and the loss function, and there is no need to store the full matrix in computer memory at any point.
Details on the number of minimization steps and how to propagate information between cones can be found in \autoref{sec:boundary-conditions}.

As a final step of our reconstruction, after performing the above described minimization, we draw posterior samples using the Fisher-Laplace approximation.
In the Fisher-Laplace approximation, the posterior distribution is approximated as
\begin{align}
    P(\xi|\mathcal{D}) = \mathcal{G}(\xi|\xi_\text{MAP},F^{-1}(\xi_\text{MAP})) \label{eq:define-fisher-laplace}
\end{align}
One can compute posterior samples in this approximation by first drawing a sample
\begin{align}
    \xi_\text{samp-inv} \curvearrowleft \mathcal{G}(\xi_\text{samp-inv}|0,F(\xi_\text{MAP})) \label{eq:metric-sample}
\end{align}
and then applying $F^{-1}$ to the inverse sample $\xi_\text{samp-inv}$ via the Conjugate Gradient algorithm \citep{hestenes1952methods} and adding the resulting residual to the maximum posteriori estimate $\xi_\text{MAP}$ that was the result of our minimization.
To compute the sample of \autoref{eq:metric-sample} we apply a matrix square-root of the metric $F$ to a vector of independent normal random numbers.
The Fisher metric $F$ is composed of a prior and a likelihood contribution. The likelihood contribution is composed of the Jacobian of the inputs to our Student-T likelihood and the Fisher metric of the Student-T distribution itself. The prior contribution is the identity matrix $\mathbb{1}$.

The matrix square root of this Fisher metric, which we apply to normal random numbers in order to obtain a sample with covariance $F$, can be computed analytically:
\begin{align}
    \sqrt{F}(\xi)^T:
    \xi^t \mapsto &=
    \begin{pmatrix}
        \xi^t \vspace{.35cm}\\
        \sqrt{\frac{\theta+1}{\left(\theta+3\right)\left(\sqrt{\sigma^2_{\text{ph}}+R^\text{var}(\rho)}\right)^2}} \frac{\partial R^\text{avg}(s)}{\partial \xi}(\xi^t) \vspace{.5cm} \\
        \sqrt{\frac{2\theta}{\left(\theta+3\right)\left(\sqrt{\sigma^2_{\text{ph}}+R^\text{var}(\rho)}\right)^2}}\frac{\partial \sqrt{\sigma^2_{\text{ph}}+R^\text{var}(\rho)}}{\partial \xi}(\xi^t)
    \end{pmatrix} \label{eq:fisher-sqrt}\\
\end{align}

\section{Results and discussion}
\label{sec:results}

\subsection{Results}
\begin{figure*}[p]
        \centering
        \begin{subfigure}[t]{.46\textwidth}
                \includegraphics[trim={0.8cm .2cm .5cm .5cm}, clip, width=.9\textwidth]{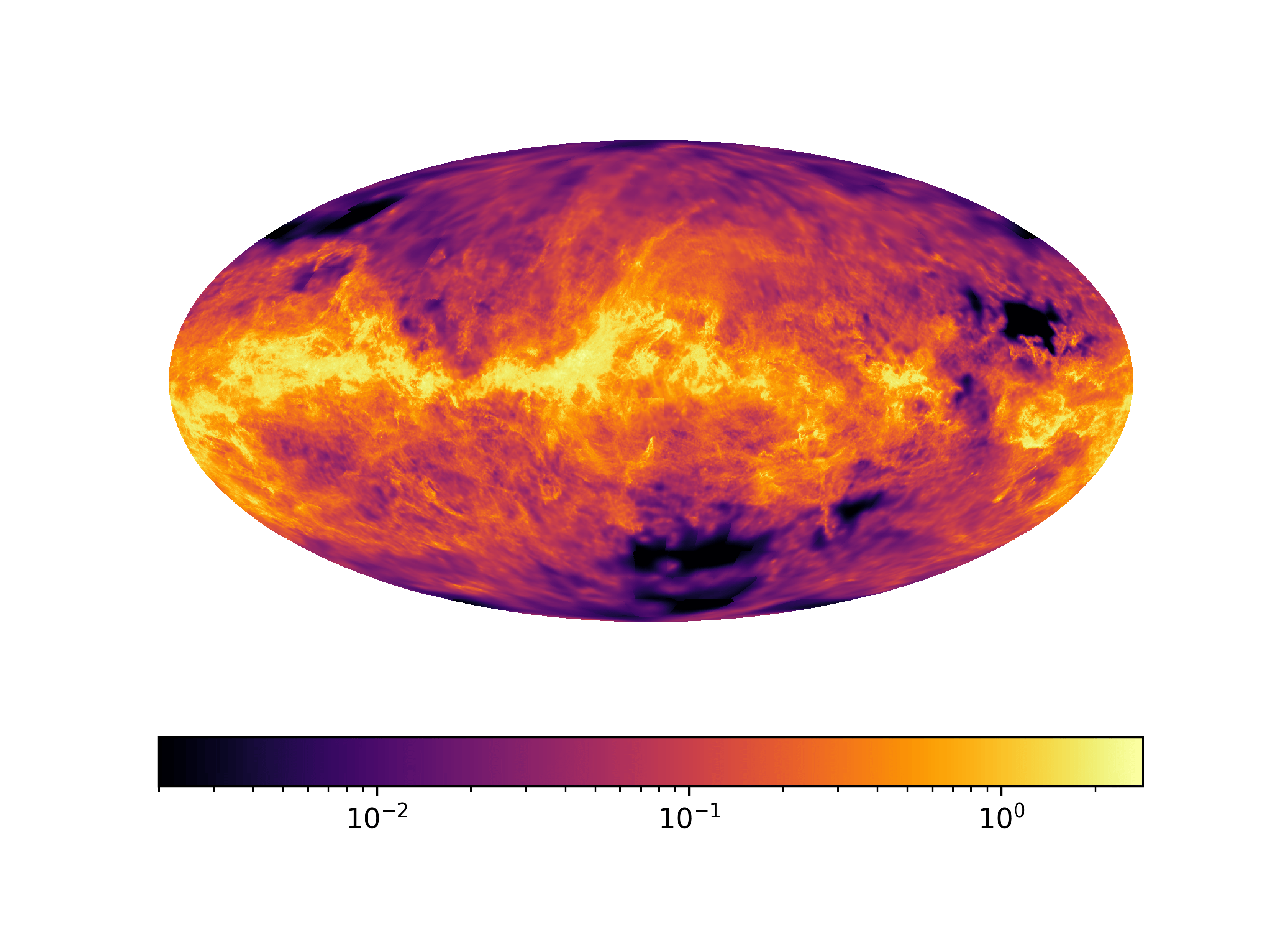}
        \caption{
        }
        \end{subfigure}
        ~
        \begin{subfigure}[t]{.46\textwidth}
                \includegraphics[trim={0.8cm .2cm .5cm .5cm}, clip, width=.9\textwidth]{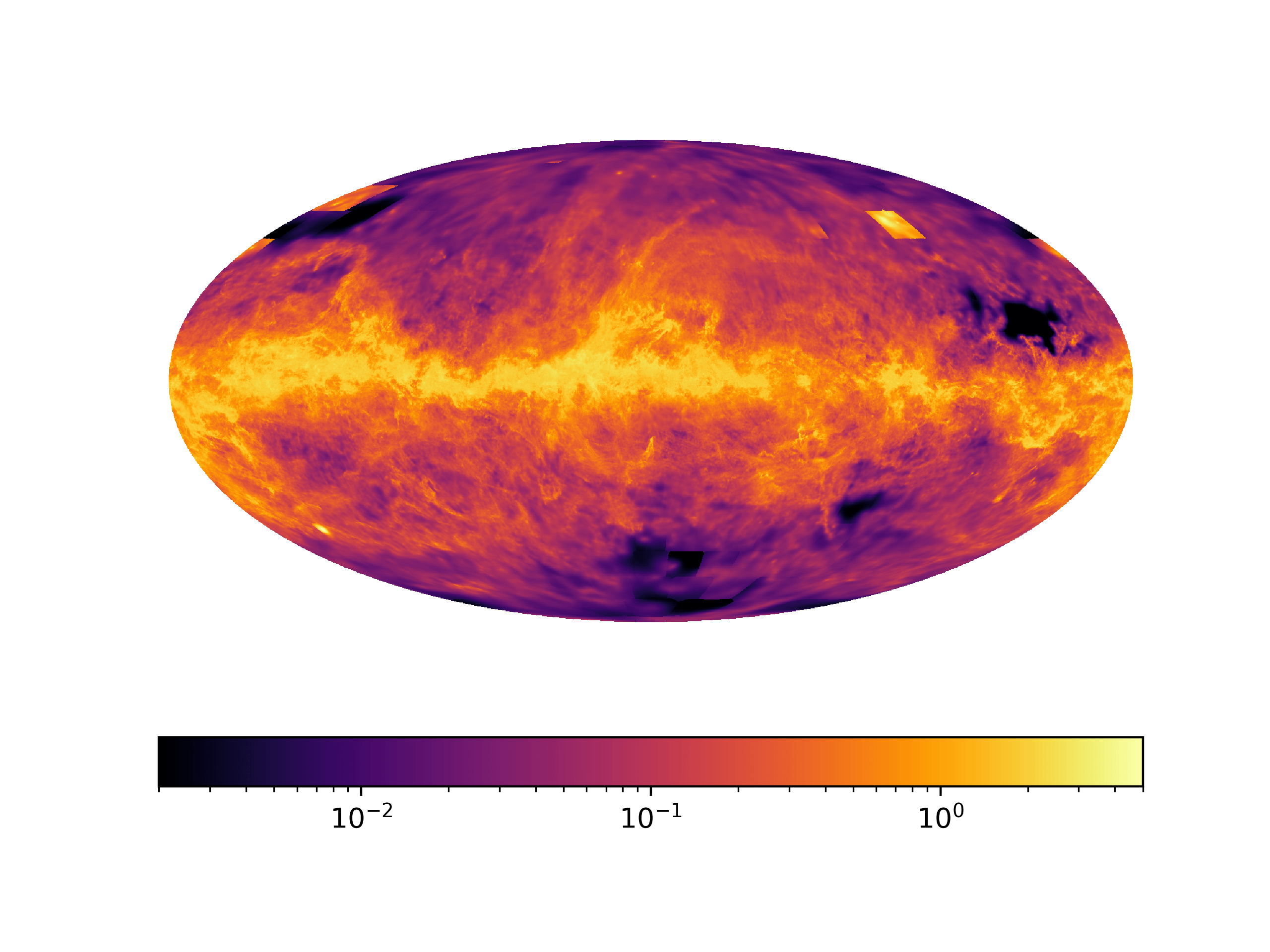}
                \caption{
}
        \end{subfigure}
        \\
        \begin{subfigure}[t]{.46\textwidth}
                \includegraphics[trim={0.8cm .2cm .5cm .5cm}, clip, width=.9\textwidth]{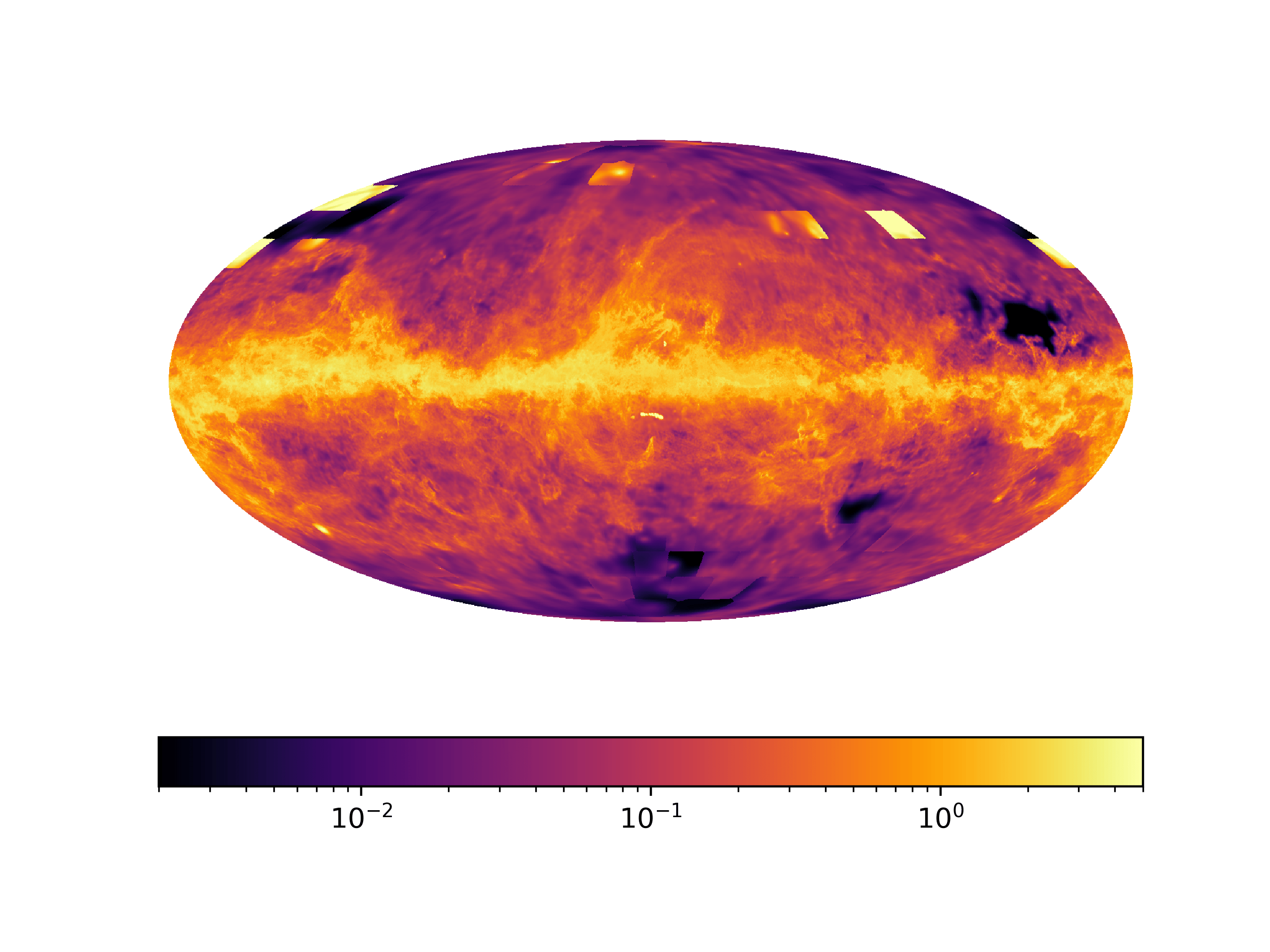}
        \caption{
        }
        \end{subfigure}
        ~
        \begin{subfigure}[t]{.46\textwidth}
                \includegraphics[trim={0.8cm .2cm .5cm .5cm}, clip, width=.9\textwidth]{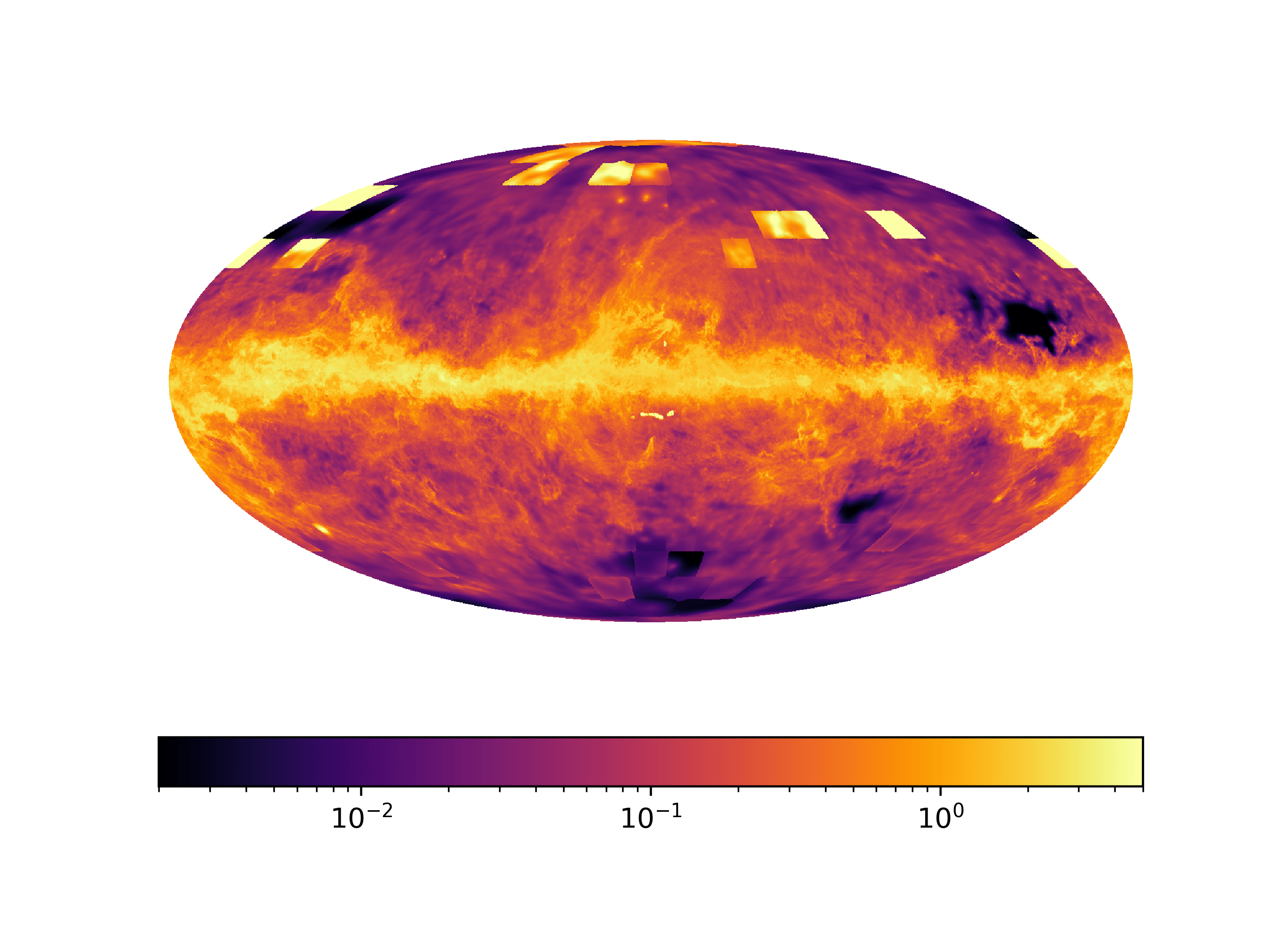}
        \caption{
                }
        \end{subfigure}
        \\
        \begin{subfigure}[t]{.46\textwidth}
                \includegraphics[trim={0.8cm .2cm .5cm .5cm}, clip, width=.9\textwidth]{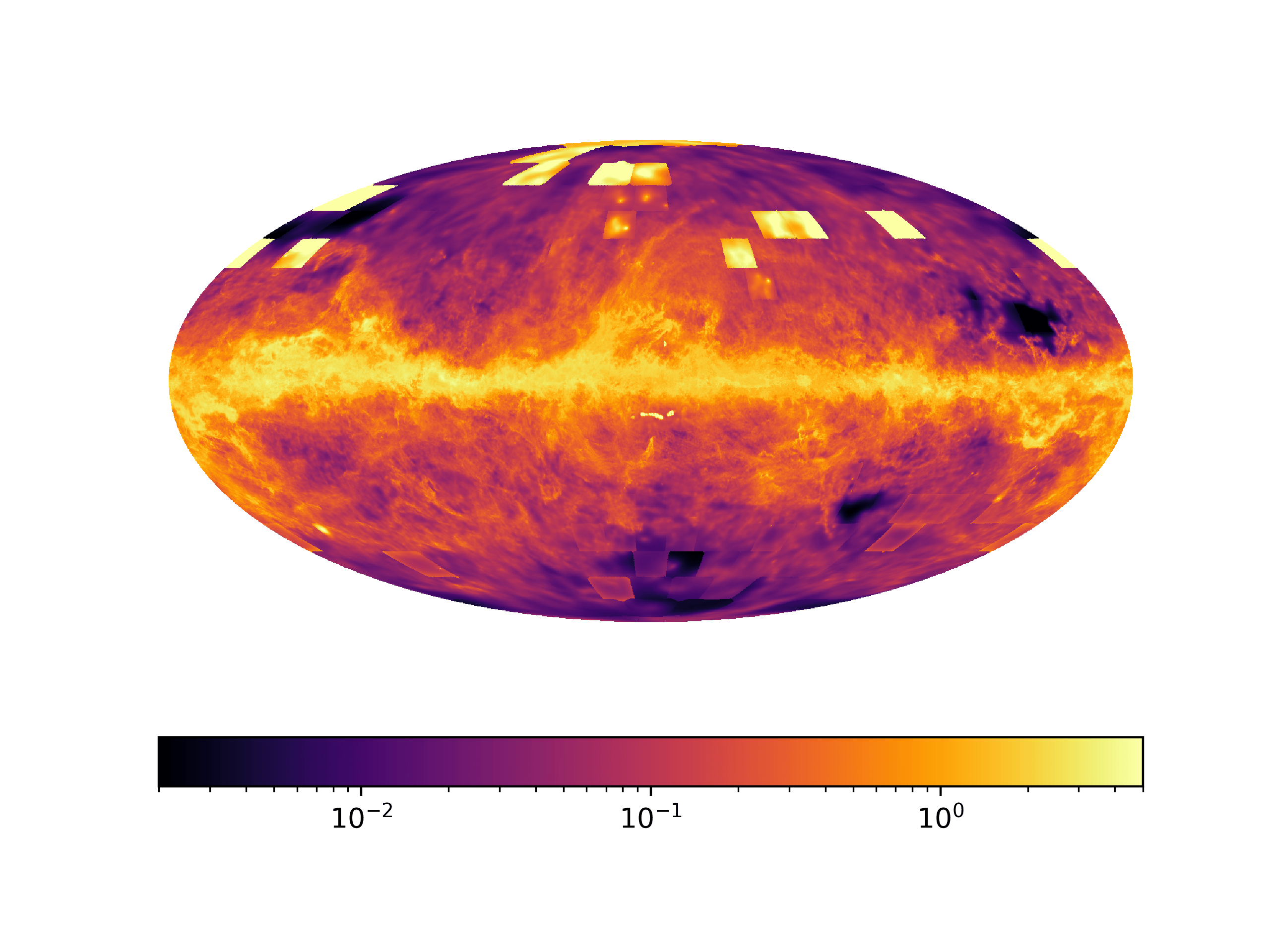}
        \caption{
        }
        \end{subfigure}
        ~
        \begin{subfigure}[t]{.46\textwidth}
                \includegraphics[trim={0.8cm .2cm .5cm .5cm}, clip, width=.9\textwidth]{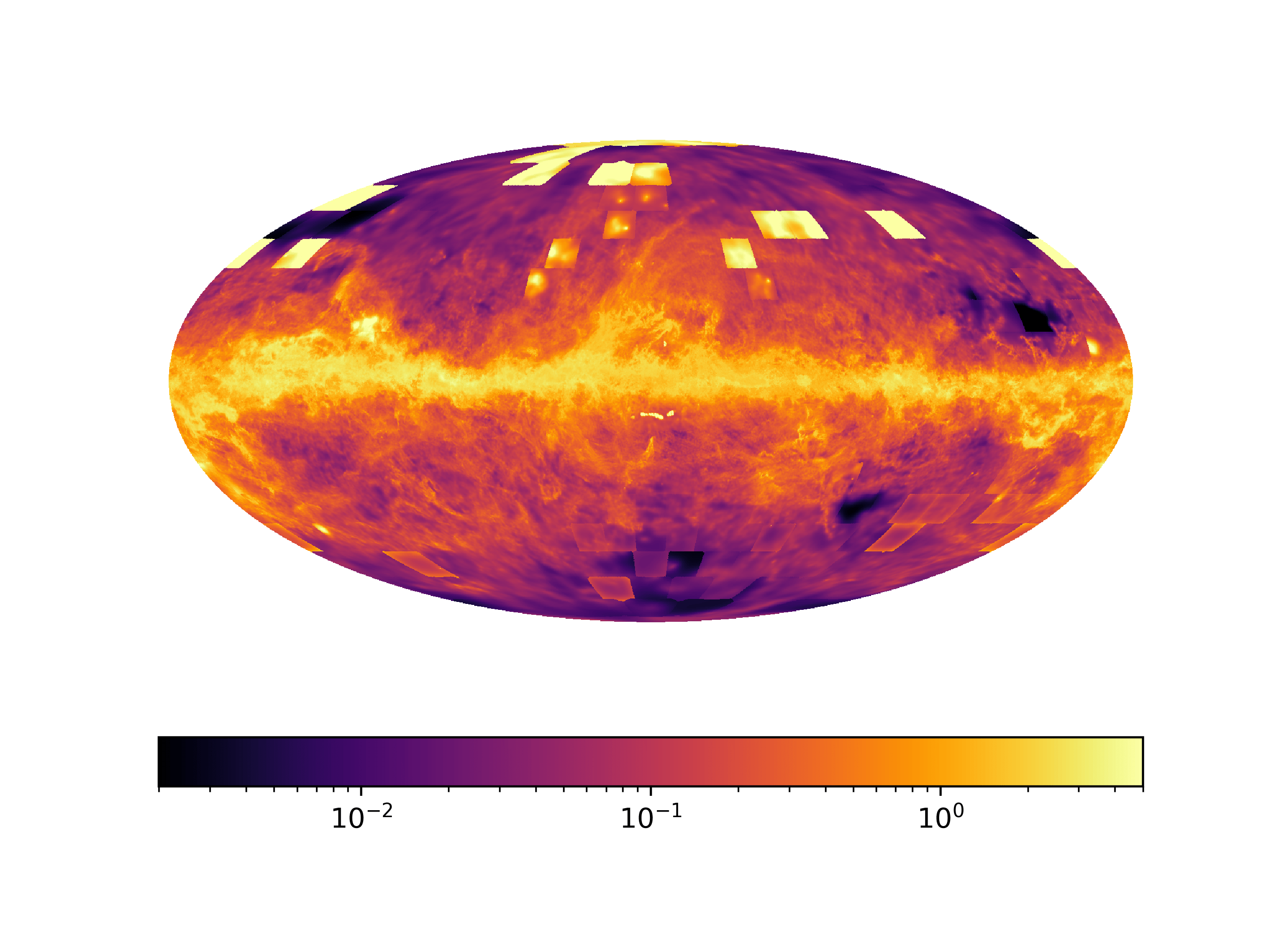}
        \caption{
        }
        \end{subfigure}
    \caption{\label{fig:hp-distance}
    Integrated extinction in magnitudes to distances in $1\,$kpc distance increments, with the top left panel showing the integrated extinction to $1\,$kpc, and the bottom right panel to distances up $6\,$kpc , where our map is more unreliable.
    All plots show the extinction in the G-band in magnitudes using a logarithmic color scale adapted to the range of parameters of the plot.
    The distorted rectangular shapes visible in the map are for cones which did not converge well beyond some distance. This is typically the case for regions with a low density of measurement points, i.e. towards the poles, where there are less stars, or behind dense foreground clouds.
        }                       %
\end{figure*}

The name of the here presented map is ICECONE (\textbf{I}nterstellar \textbf{C}one-based \textbf{E}stimate of \textbf{CO}rrelated \textbf{N}on-negative \textbf{E}xtinction).
The resolution elements of the map cover an area of $16\,\text{kpc}$ radius,
but the data only allow for a meaningful reconstruction in a subvolume.
Fig.\,\ref{fig:hp-distance} shows integrated extinction to different distance threshold from $1$ to $6\,\text{kpc}$.

\begin{figure*}[p]
        \centering
        \begin{subfigure}[t]{.46\textwidth}
                \includegraphics[trim={0.8cm .2cm 2.0cm 1cm}, clip, width=.95\textwidth]{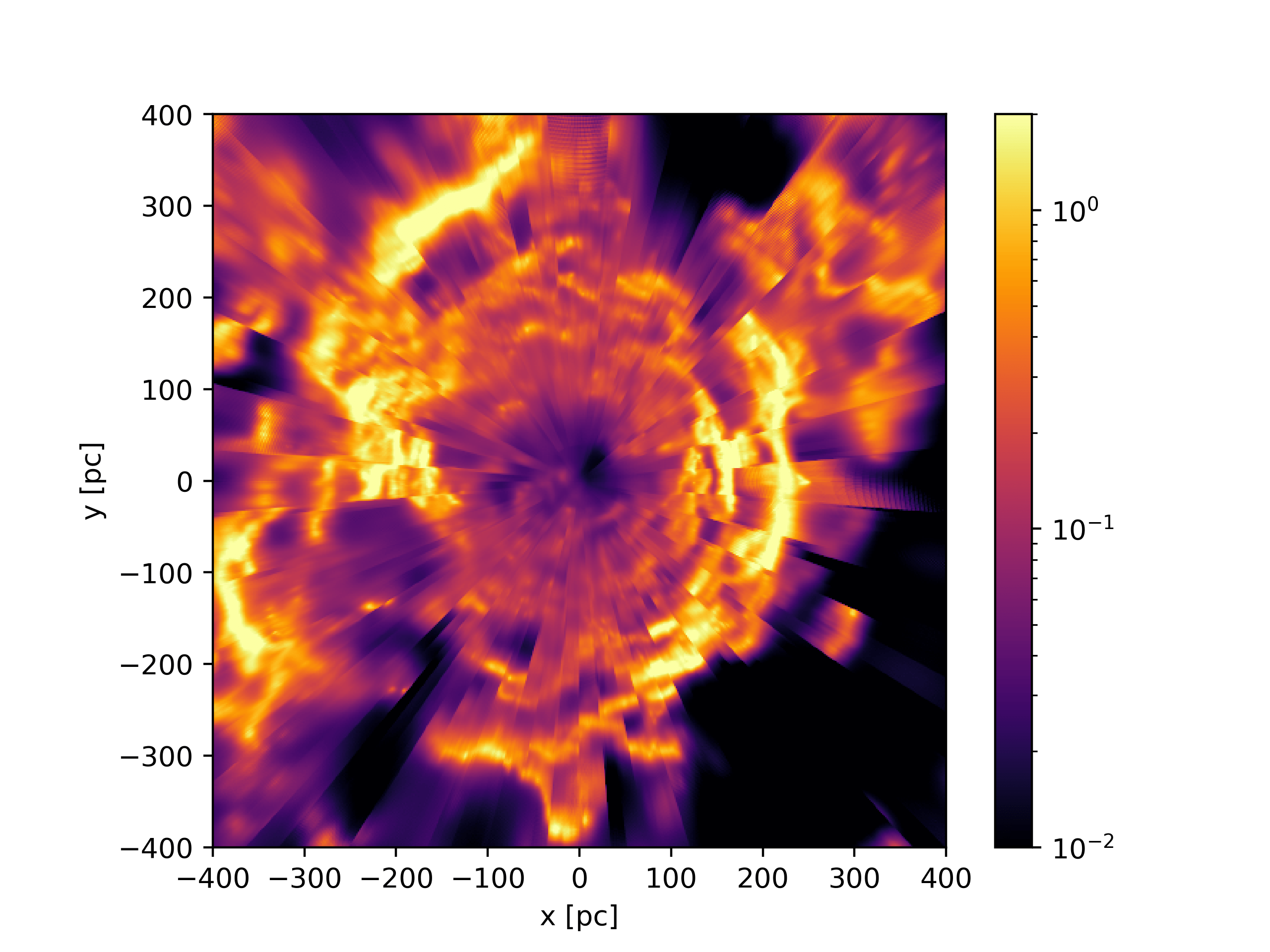}
        \caption{
        }
        \end{subfigure}
        ~
        \begin{subfigure}[t]{.46\textwidth}
                \includegraphics[trim={0.8cm .2cm 2.0cm 1cm}, clip, width=.95\textwidth]{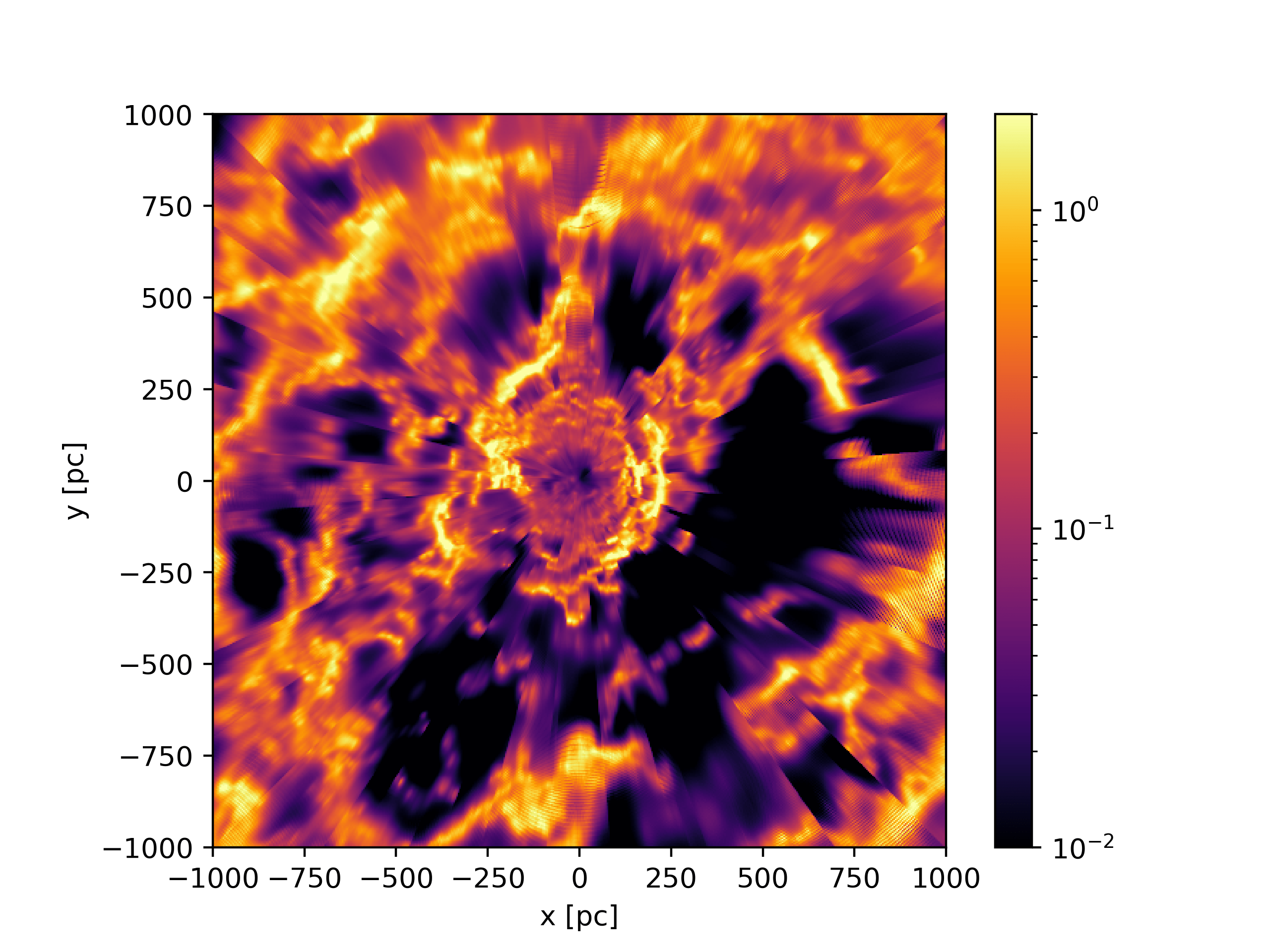}
                \caption{
}
        \end{subfigure}
        \\
        \begin{subfigure}[t]{.46\textwidth}
                \includegraphics[trim={0.8cm .2cm 2.0cm 1cm}, clip, width=.95\textwidth]{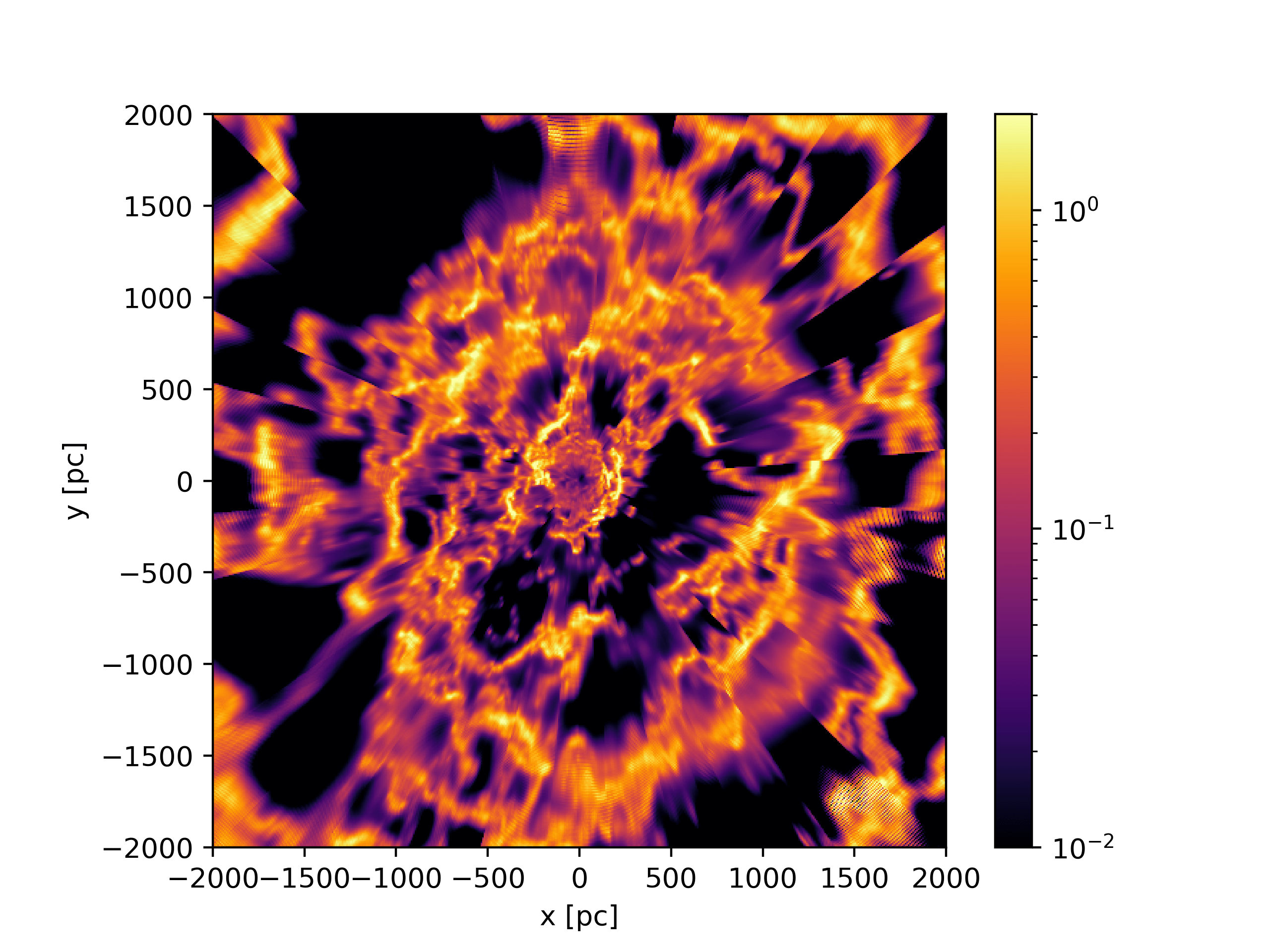}
        \caption{
        }
        \end{subfigure}
        ~
        \begin{subfigure}[t]{.46\textwidth}
                \includegraphics[trim={0.8cm .2cm 2.0cm 1cm}, clip, width=.95\textwidth]{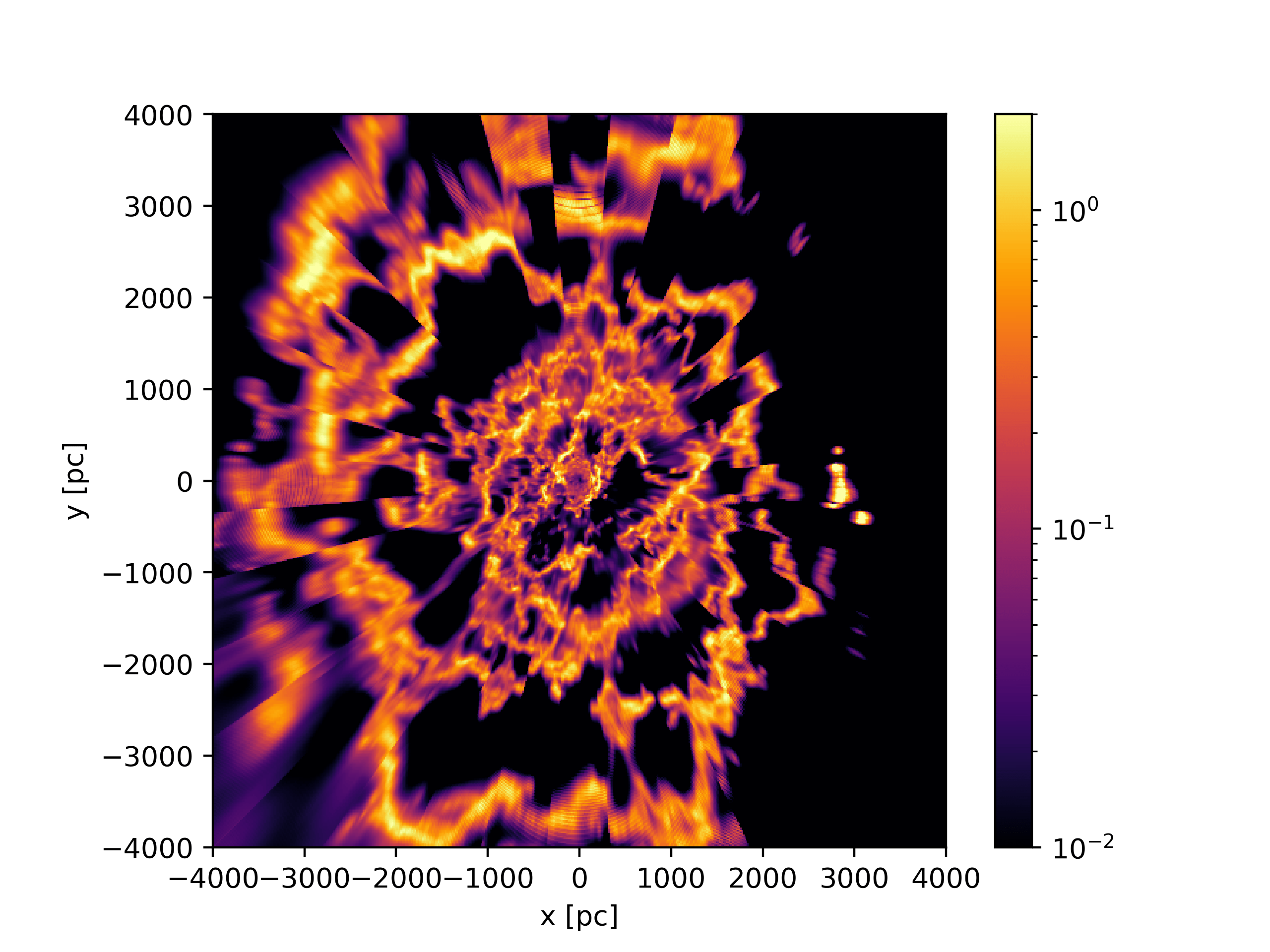}
        \caption{
                }
        \end{subfigure}
        \\
        \begin{subfigure}[t]{.46\textwidth}
                \includegraphics[trim={0.8cm .2cm 2.0cm 1cm}, clip, width=.95\textwidth]{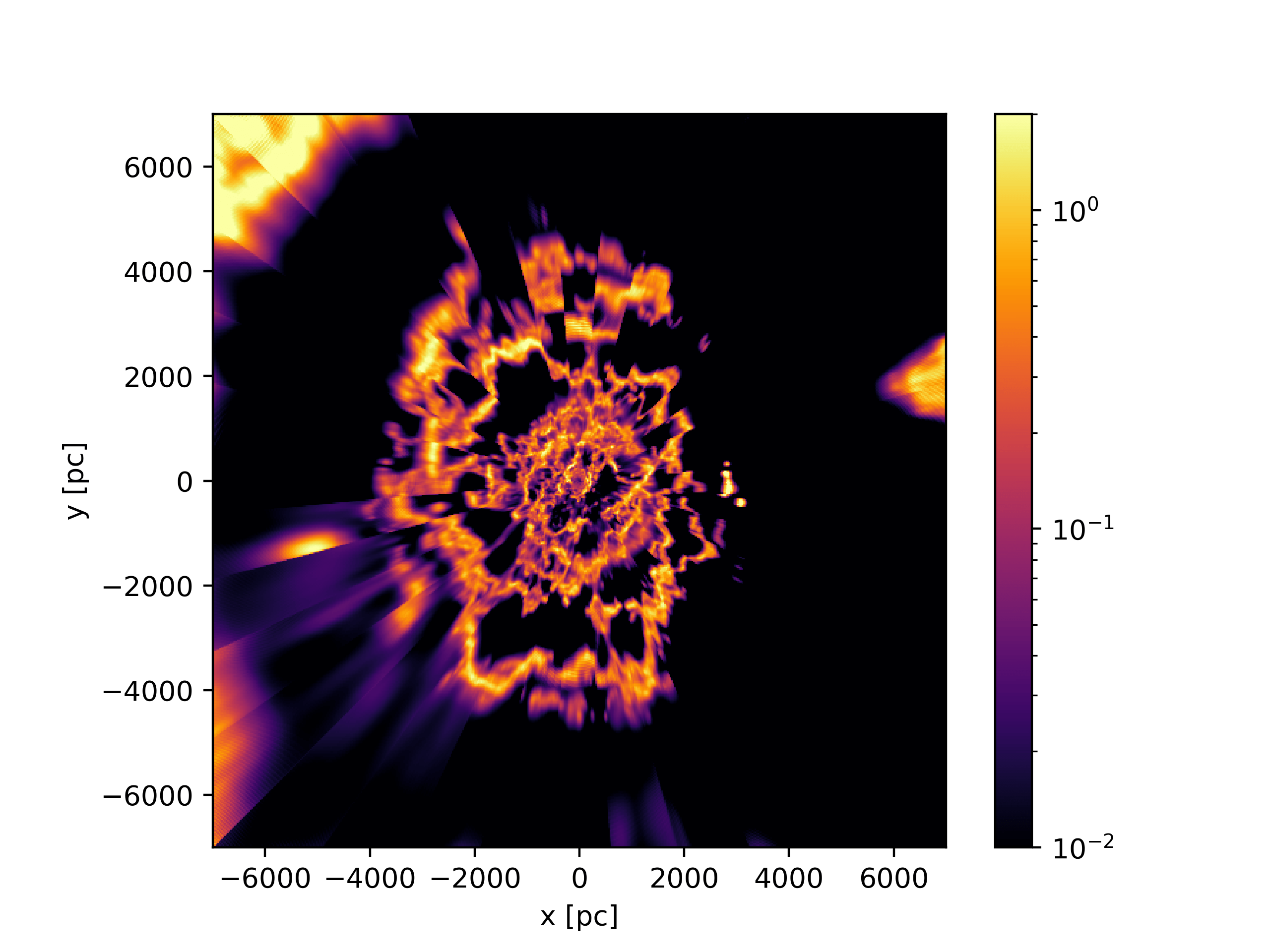}
        \caption{
        }
        \end{subfigure}
        ~
        \begin{subfigure}[t]{.46\textwidth}
                \includegraphics[trim={0.6cm .2cm 2.0cm 1cm}, clip, width=.95\textwidth]{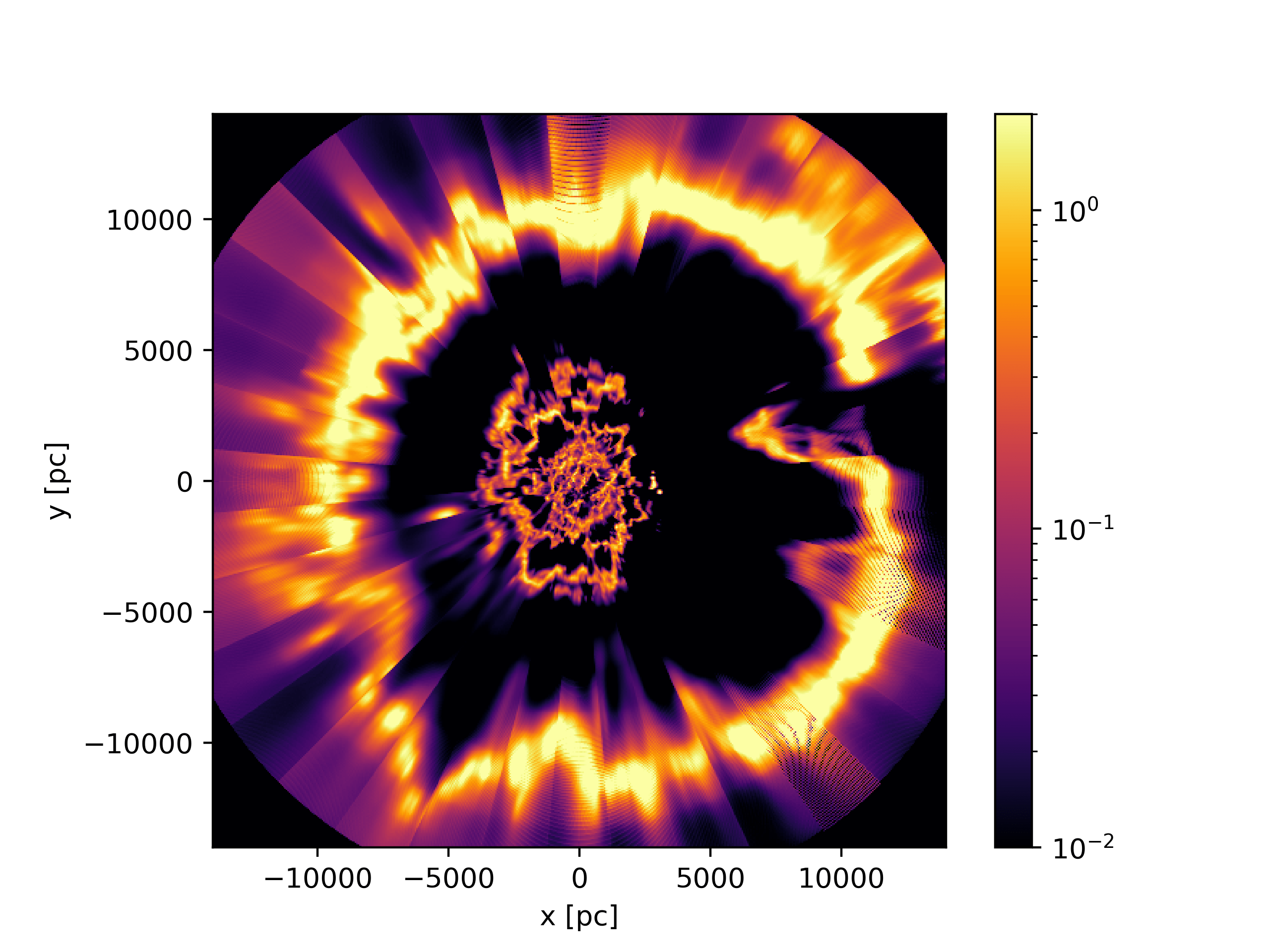}
        \caption{
        }
        \end{subfigure}
        \caption{
        \label{fig:plane-zoom}
        Integrated extinction in the Galactic plane, shown at increasing length scales.
        The Sun is at the origin, and the Galactic center is located on the right at $x=8200\,\text{pc}$, $y=0\,\text{pc}$, and $z=-11\,\text{pc}$.
        The $x$- and $y-$ axis are Galactic coordinates, where the dust extinction density was integrated over Galactic $z$- axis.
        The additional regions shown by the bottom two panels are outside the trusted region of the map, where due to data scarcity no structures could be reconstructed reliably.
        }                       %
\end{figure*}

Figure \ref{fig:plane-zoom} shows dust extinction column density in the Galactic plane in distance ranges from $800\,\text{pc}$ to $16\,\text{kpc}$.
At a distance of about $4\,\text{kpc}$ the reconstruction shows a large void, which ends in a large ring of dust at $8\,\text{kpc}$.
We believe this effect to be an artifact of our reconstruction, as it is an unphysical heliocentric structure and there are not sufficiently many datapoints to yield a robust reconstruction at these distances.
Realistic filamentary dust structures can be seen up to a distance of $6\,\text{kpc}$, with the precision of the reconstruction being better for nearby dust structures.
Furthermore Figure \ref{fig:plane-zoom} exhibits that there are discontinuities at the boundaries of the individual angular patches, where despite our efforts to propagate information between the angular cones the reconstruction does not agree on a coherent structure.

\subsection{Validation}
\label{subsec:validation}

Compared to our previous Cartesian model in~\cite{leike2020resolving} we use iterative grid refinement~\citep{edenhofer2021sparse} to represent the spatially correlated dust extinction.
With this choice of spherical basis we are able to reconstruct volumes of a size and with a resolution that would not be computationally feasible to be represented in Cartesian coordinates.
We validate this more efficient representation against our Cartesian model.

We use the generative nature of our model in~\cite{leike2020resolving} to create a synthetic sample of dust extinction on a Cartesian grid encompassing a dust cone of $3\,\text{kpc}$ in length.
Due to the much more expensive scaling with volume of our Cartesian model, we were not able to synthetically generated significantly larger cones.
Using the Gaia star positions and synthetically sampled parallaxes, we construct a synthetic data sample from our Cartesian model:
\begin{align}
	\rho^* =\ &\rho_0 \cdot \exp{(\tau^*)} \quad\text{with}\ \tau^*\curvearrowleft \mathcal{G}(\tau^*|0,C)\ ,
	\\ A^*_i =\ & \int_0^{1/\omega^*_i}{\mathrm{d}r^\prime}\,\rho^*(r^\prime \phi_i) \quad\text{with}\ \omega^*_i \curvearrowleft \mathcal{G}(\omega^*_i|\omega_i,\sigma_{\omega,i})\ ,
\end{align}
whereby an asterisk denotes the synthetic nature of the quantity to which it is attached.
We add synthetic noise according to the noise statistics described in~\autoref{table:good-photoflags}
\begin{equation}
	a^*_i \curvearrowleft T(a^*_i|A^*_i + m_{\mathrm{ph}(i)}, \sigma_{\mathrm{ph}(i)}, \nu_{\mathrm{ph}(i)})\ .
\end{equation}
Note, since we synthetically sample the ``true'' distance, we do not need to account for any supplementary noise due to the parallax uncertainty in the data nor do we need to account for the parallax uncertainty in the synthetic extinction.
Lastly, we reconstruct the dust extinction from the artificially generated extinction data $D=(a^*_i)_i$ using our proposed model.

\begin{figure}
		\includegraphics[trim={0.2cm .2cm 1.0cm 0cm}, clip, width=.47\textwidth]{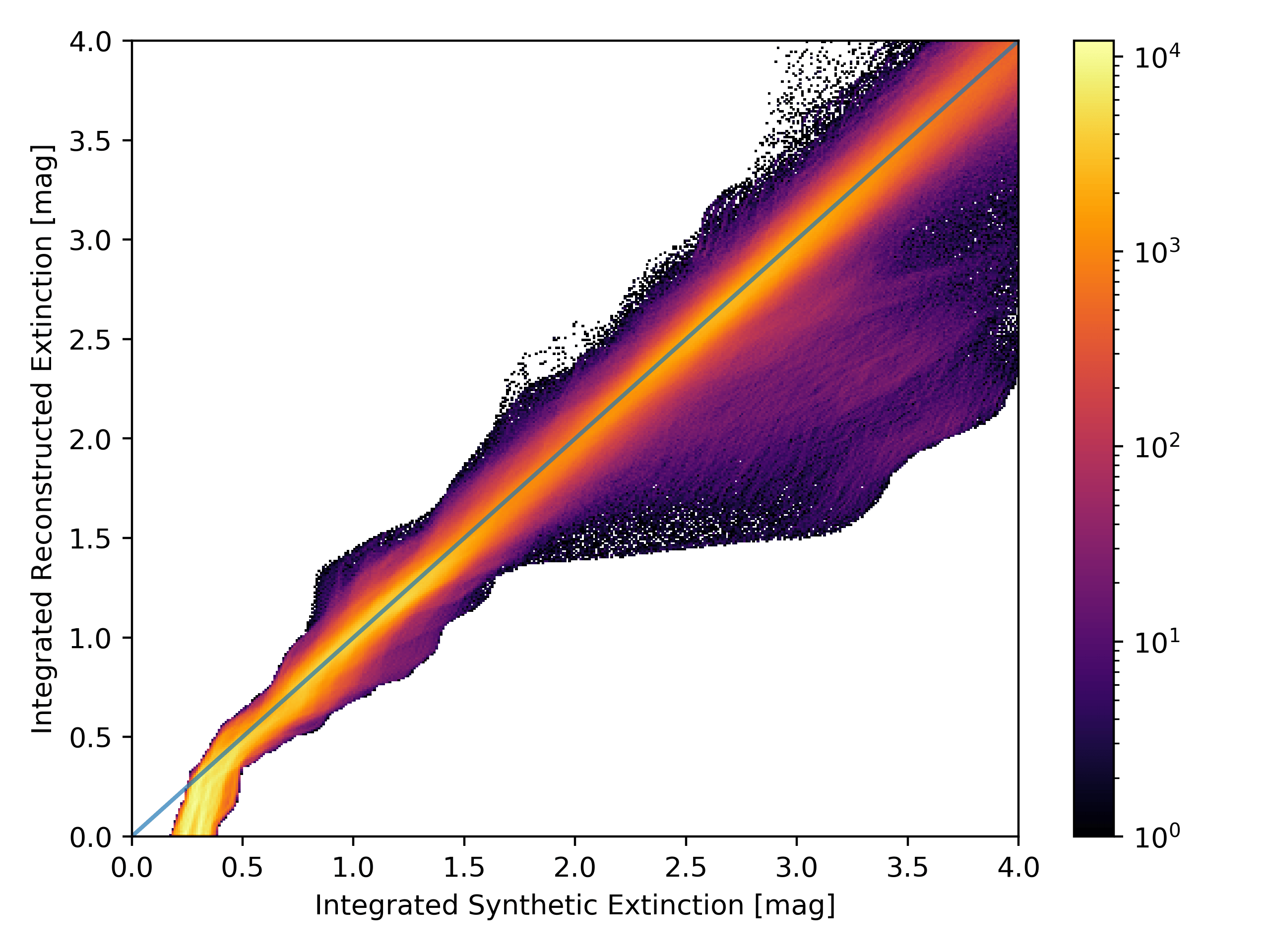}
        \caption{\label{fig:validation_heatmaps}
			Integrated extinction up to $4\,\text{mag}$ of the dust reconstruction based on synthetic data compared to the synthetic truth. The bisector is shown in blue.
            Stronger deviations in the first $0.5\,\text{mag}$ are a consequence of the ground truth containing dust at the inner $40\,\text{pc}$, which cannot be represented by the reconstruction as its coordinate grid only starts at $40\,\text{pc}$.
        }                       %
\end{figure}

\autoref{fig:validation_heatmaps} depicts a heatmap of the reconstructed versus the synthetic integrated extinction cut at $4\,\text{mag}$.
Except for the first $0.25\,\text{mag}$ the reconstruction versus the synthetic truth follows the bisector.
At $0.25\,\text{mag}$ of the synthetic truth, the reconstruction increases steeply from zero to the level of the truth.
With increasing extinction the correspondence degrades again and the reconstruction deviates more strongly from the synthetic truth.
The variability of the reconstruction increases with magnitude.

Overall the reconstruction is in agreement with the synthetic truth.
The discrepancy at $0.25\,\text{mag}$ is due to our definition of the grid described in \autoref{sec:grid-refinement-for-cones}.
Our synthetic sample features dust in the first $40\,\text{pc}$.
This dust cannot be represented by our logarithmically spaced radial bins, as the first voxel begins at $40\,\text{pc}$.
In the synthetic truth the innermost voxels absorb the integrated extinction of the first $40\,\text{pc}$ while the reconstruction needs a couple of inner voxels to catch up without violating the prior too much.
We do not believe this to have an effect on the actual reconstruction, as the region around the Sun is known to be sufficiently void of dust \citep{leike2020resolving}.

\subsection{Discussion}
\label{sec:discussion}

We explore different ways to assess the consistency of our dust reconstruction with findings in the literature.
By exploiting correlations, we are able to predict precise distance estimates for all larger nearby dust clouds.
This is probably the most unique aspect of our reconstruction, which allows to validate it via independent measurements of dust cloud distances.
Such exist in the form of masers for which parallaxes can be measured precisely through VLBI observations.
Masers are associated with young star forming regions typically embedded in large dust overdensities and therefore provide validation data for the distances of dust clouds resulting from our reconstruction.

\begin{figure}[ht]
    \centering
                \includegraphics[trim={0.5cm .0cm 1.5cm .6cm}, clip, width=.5\textwidth]{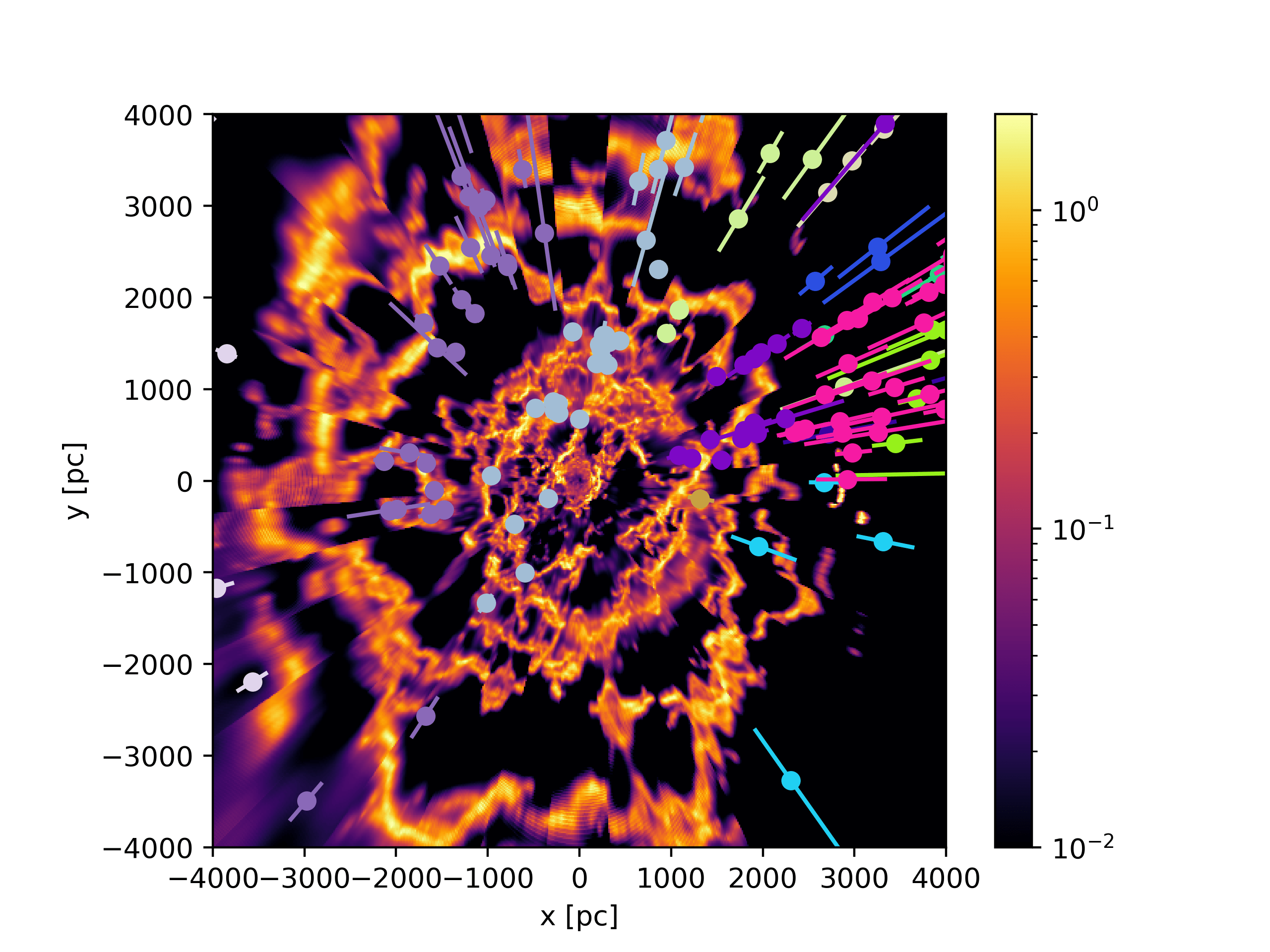}
        \caption{
        \label{fig:maser-plane}
        Galactic dust extinction in the Galactic plane together with masers of \cite{reid2019trigonometric}.
        The dust is shown on logarithmic scale in magnitudes of G-band extinction.
        Masers associated to different spiral arms are shown in different colors.
        The errorbars show one sigma of their radial distance uncertainty.
        }                       %
\end{figure}

In \autoref{fig:maser-plane} we show an $8\,\text{kpc}\times8\,\text{kpc}$ slice through our dust density with the masers of \citet{reid2019trigonometric} plotted on top.
One can see a nice correspondence for more nearby masers in this projection up to radii of  $4\,\text{kpc}$. However, larger deviations between maser locations and dust clouds are found towards the Galactic center and at larger distances indicating that our dust cloud distances should be less trustworthy there.

\begin{figure}[ht]
    \centering
                \includegraphics[trim={0.0cm .0cm 0cm .0cm}, clip, width=.5\textwidth]{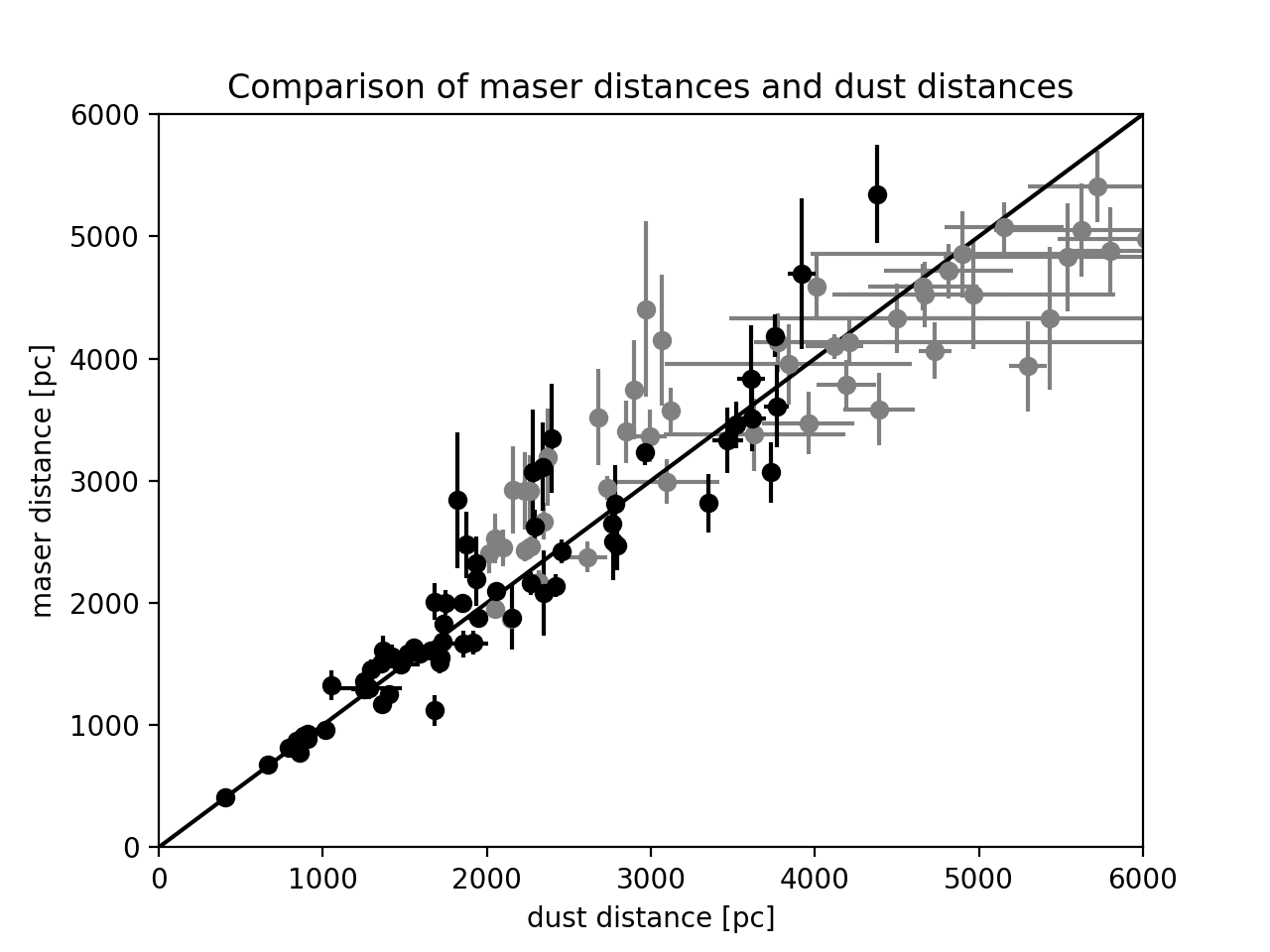}
        \caption{
        \label{fig:maser-distance}
        Distance estimates of masers according to our dust map (x-axis) and according to \cite{reid2019trigonometric} (y-axis).
        The corresponding errorbars show one standard deviation of uncertainty.
        For this analysis we only use masers that are within $6\,\text{kpc}$ with $84\,\%$ probability and which have a relative distance uncertainty smaller than $20\%$.
        Points shown in grey are masers that have no clear counterpart in our dust map, i.e. our dust map does not exceed a differential extinction of $10^{-4}\,\nicefrac{\text{mag}}{\text{pc}}$ over a distance range of three standard deviations of maser distances.
        See \autoref{sec:discussion} for more details how we assign the distance estimates given out map.
        }                       %
\end{figure}

To avoid biases arising from projection effects, we also directly probe the LOS-density for LOSs containing masers.
In \autoref{fig:maser-distance} we show distances towards nearby masers of \citet{reid2019trigonometric} according to our dust map.
Hereby we get our distance estimate by searching a distance range $r$ of
\begin{align}
    \frac{1}{\omega+3\sigma_\omega} < r < \frac{1}{\omega-3\sigma_\omega}\ ,
\end{align}
where $\omega$ is the masers parallax and $\sigma_\omega$ is the parallax uncertainty.
We consider all regions which are within one $e$-fold of the maximal dust density along that distance range as potential suitable host clouds.
Our mean and standard deviation for the maser position are then computed by considering all suitable locations equally likely.
Note that through this procedure our maser distance estimate never deviates more than three standard deviations from the estimate of \citet{reid2019trigonometric}.
If the maximal density along the line of sight does not exceed a differential extinction of $10^{-4}\,\nicefrac{\text{mag}}{\text{pc}}$ over the considered distance range, then we say that our map is not compatible with that maser, marked with the grey dots in \autoref{fig:maser-distance}.
Up to $2\,\text{kpc}$ our dust map is compatible with all masers, with reasonably agreeing distance estimates.
From $2\,\text{kpc}$ to $4\,\text{kpc}$ our dust map is compatible with a large fraction of the masers in that distance interval, but is in generally less reliable.
From $4\,\text{kpc}$ onward there is only one compatible maser, which might be coincidental, and we conclude that our dust map is not reliable at these distances.

\begin{figure*}
        \begin{subfigure}[t]{.46\textwidth}
                \includegraphics[trim={0.8cm .2cm 2.0cm 1cm}, clip, width=.95\textwidth]{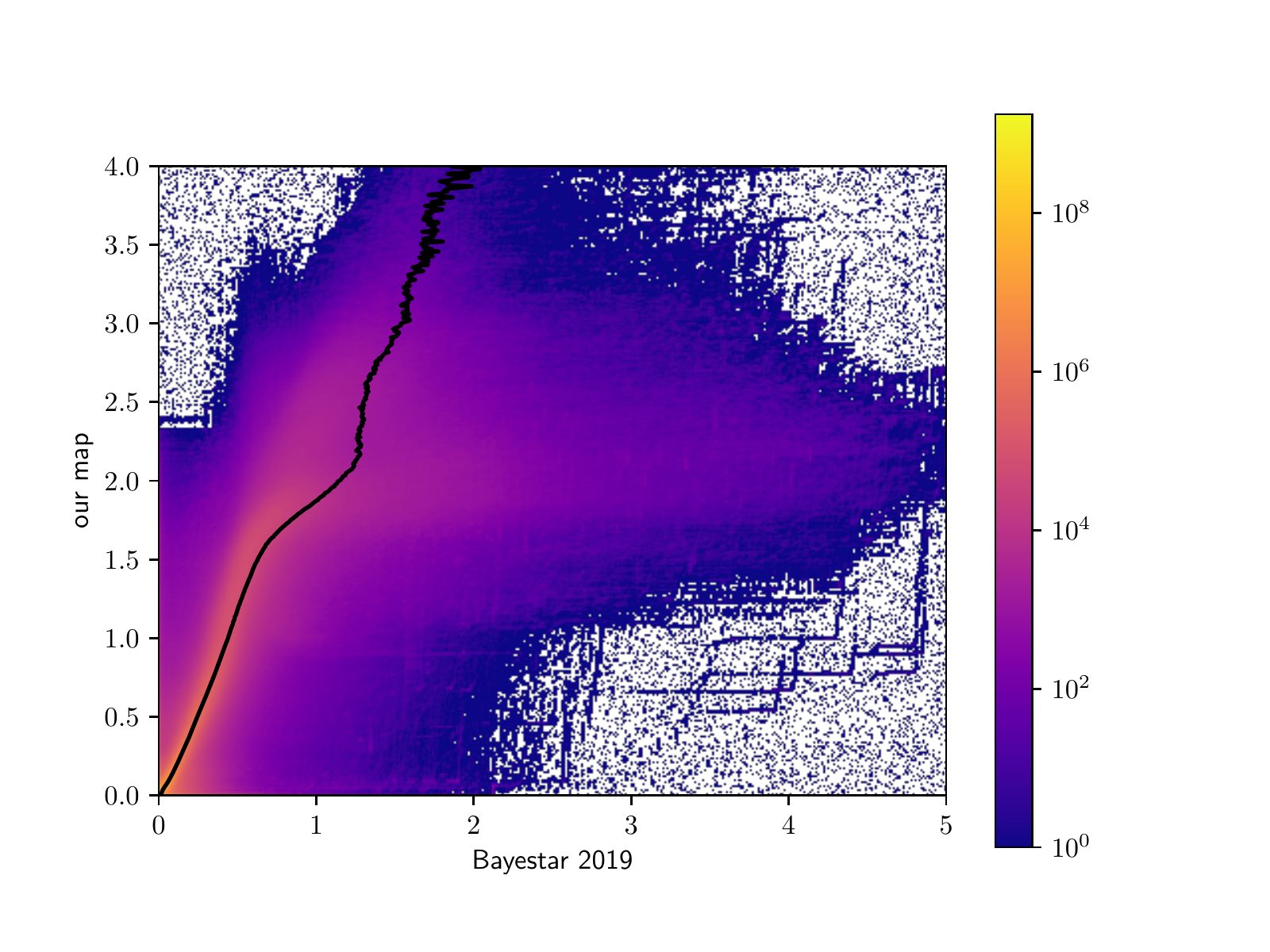}
        \caption{
        }
        \end{subfigure}~
        \begin{subfigure}[t]{.46\textwidth}
                \includegraphics[trim={0.8cm .2cm 2.0cm 1cm}, clip, width=.95\textwidth]{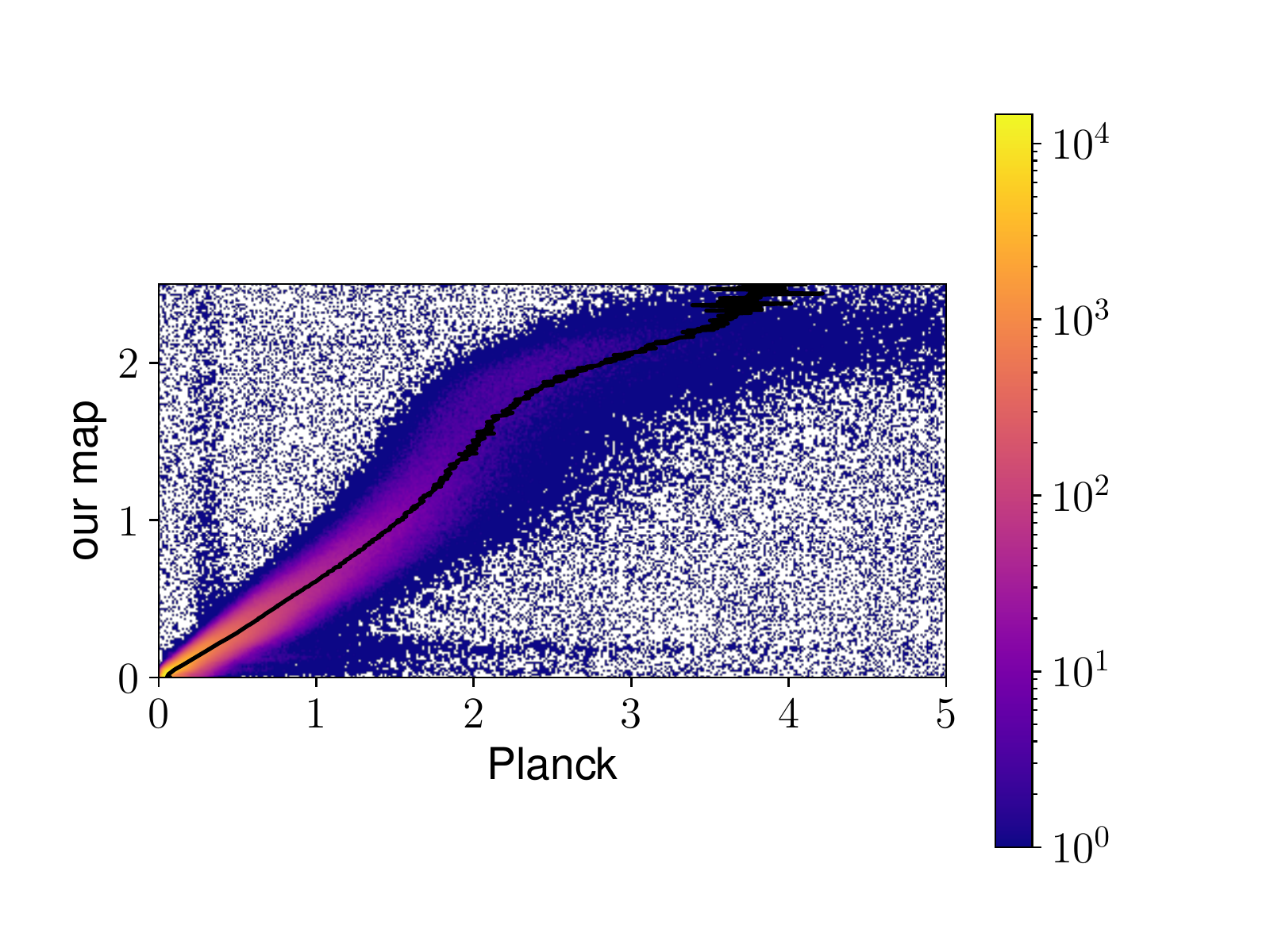}
        \caption{
        }
        \end{subfigure}
        \caption{
        \label{fig:heatmaps}
        Comparison of integrated extinction for different dust maps.
        The left panel shows counts of pairs of integrated extinction of our map and Bayesstar2019 for every pixel within $-4\,\text{kpc}<x<2\,\text{kpc}$, $-4\,\text{kpc}<y<4\,\text{kpc}$, and $-1\,\text{kpc}<z<1\,\text{kpc}$.
        The right panel shows pairs of integrated extinction of our map at a distance of $2\,\text{kpc}$ and the Planck Draine\&Li $A_V$ dust model, renormalized to match quasar extinction.
        We multiplied our map with $1.202$ to convert G-band extinction $A_G$ into visual extinction $A_V$.
        }                       %
\end{figure*}

\begin{figure*}
        \begin{subfigure}[t]{.46\textwidth}
                \includegraphics[trim={0.8cm .2cm 2.0cm 1cm}, clip, width=.95\textwidth]{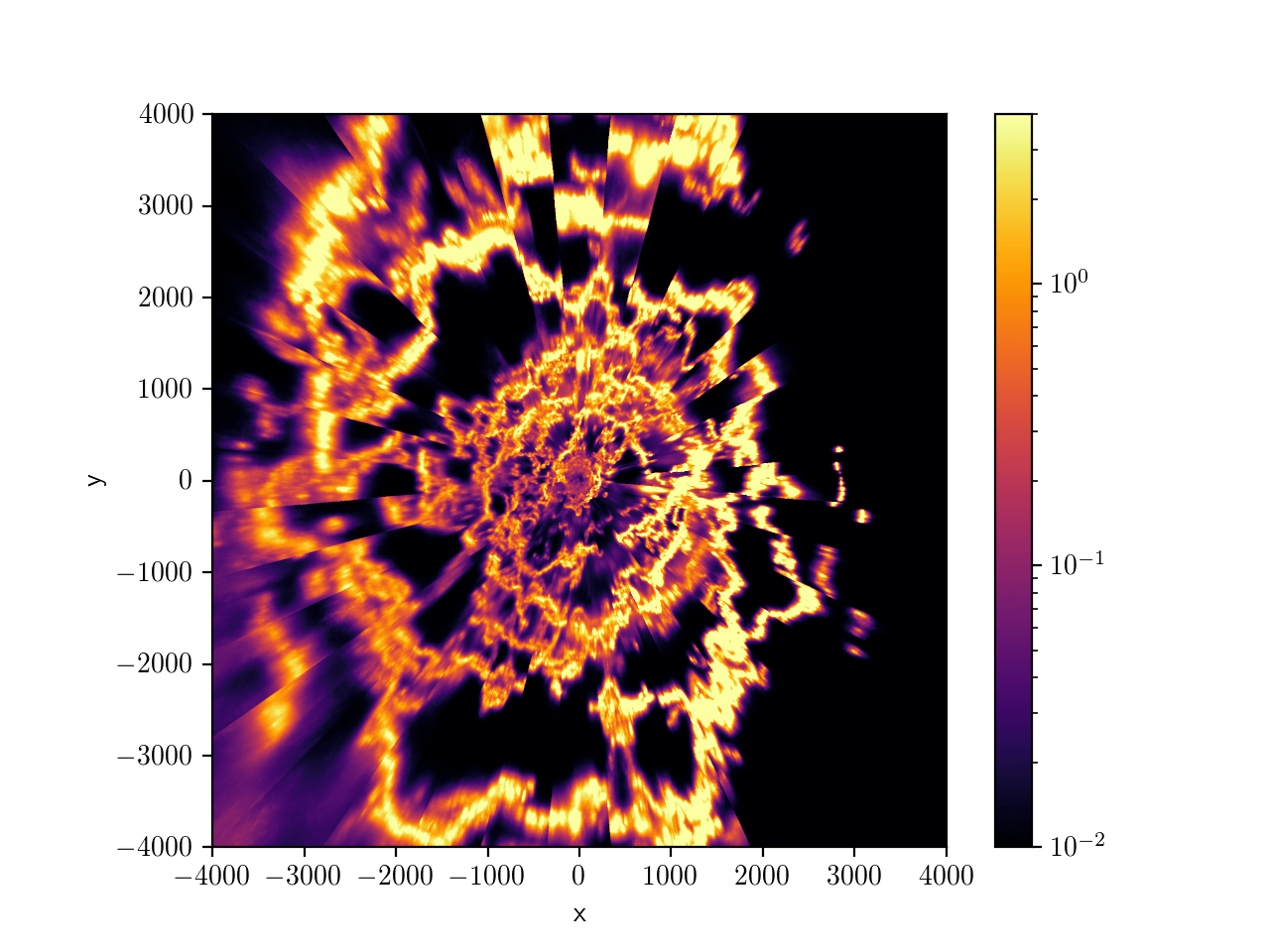}
        \caption{
        }
        \end{subfigure}~
        \begin{subfigure}[t]{.46\textwidth}
                \includegraphics[trim={0.8cm .2cm 2.0cm 1cm}, clip, width=.95\textwidth]{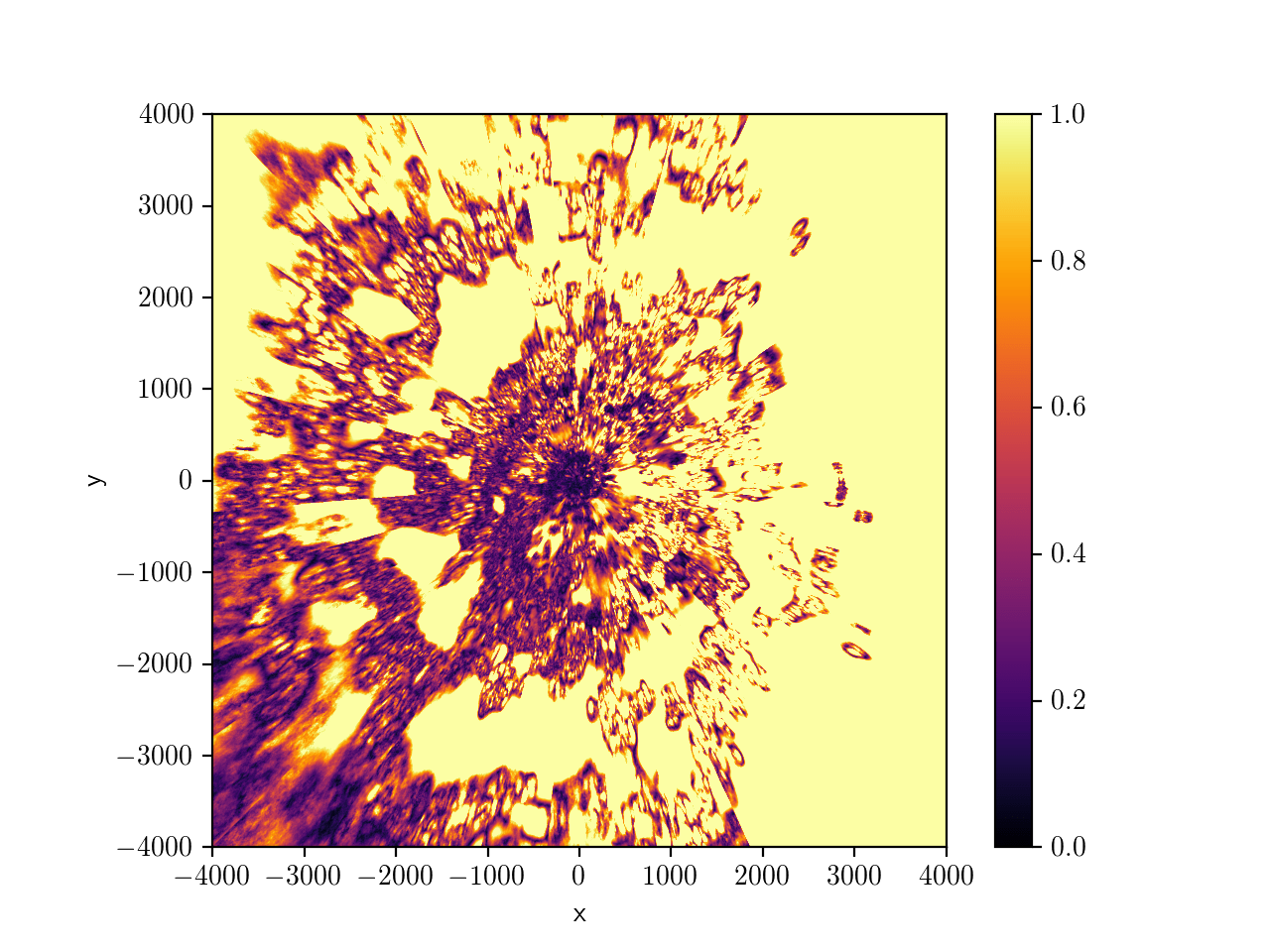}
        \caption{
        }
        \end{subfigure}\\
        \begin{subfigure}[t]{.46\textwidth}
                \includegraphics[trim={0.8cm .2cm 2.0cm 1cm}, clip, width=.95\textwidth]{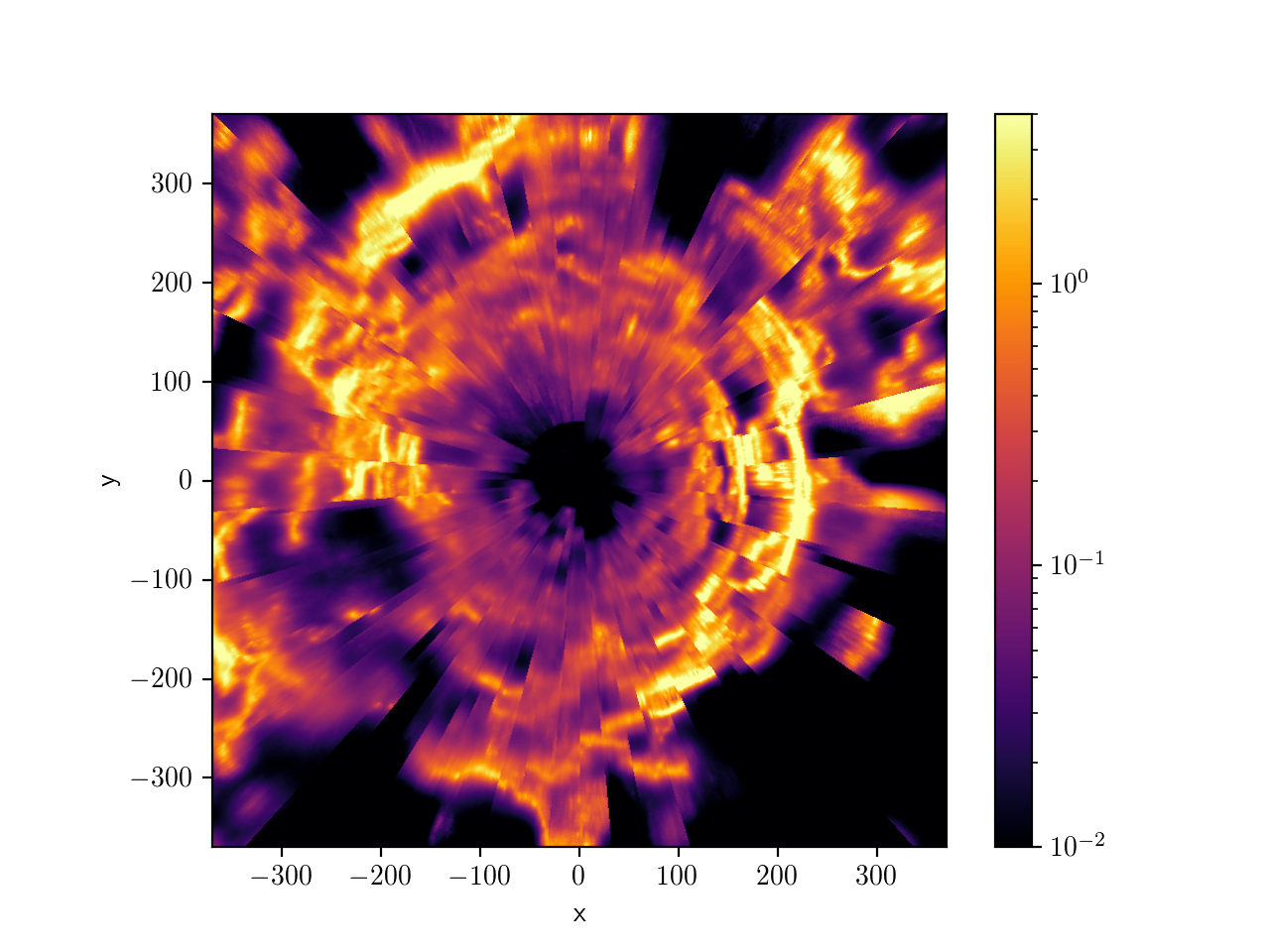}
        \caption{
        }
        \end{subfigure}~
        \begin{subfigure}[t]{.46\textwidth}
                \includegraphics[trim={0.8cm .2cm 2.0cm 1cm}, clip, width=.95\textwidth]{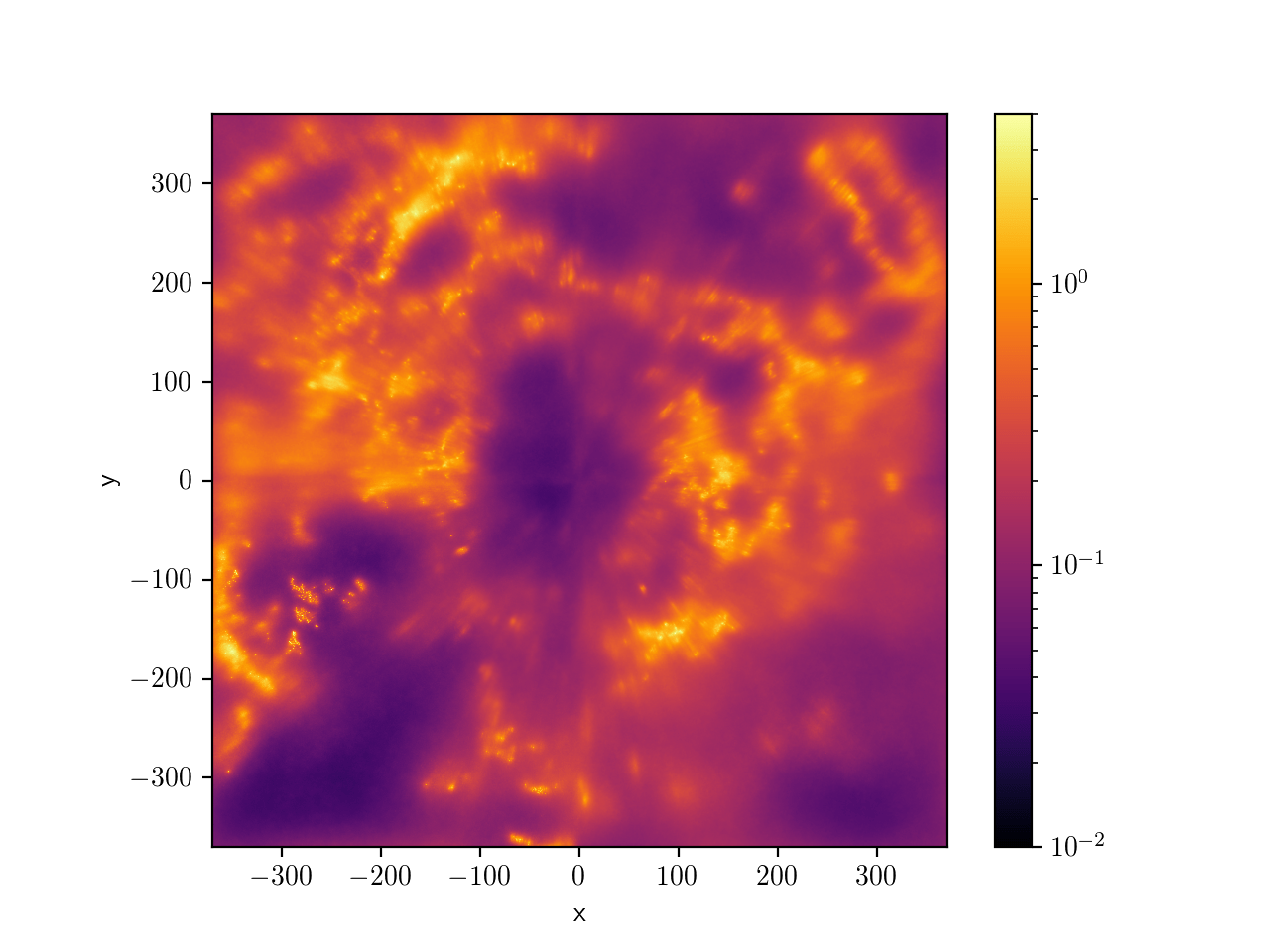}
        \caption{
        }
        \end{subfigure}
        \caption{
        \label{fig:standard-deviation-comparison}
        The left panels shows integrated extinction in the Galactic plane in a $8\,\text{kpc}\times 8\,\text{kpc}$ area (top) and a $0.8\,\text{kpc}\times 0.8\,\text{kpc}$ area (bottom).
        The $x$- and $y-$ axis are Galactic coordinates, where the dust extinction density was integrated over Galactic $z$- axis.
        The top right panels show the corresponding relative dust density uncertainties, clipped to the range from zero to one.
        The bottom right panel shows the reconstruction result of \citet{leike2020resolving} for comparison to the bottom left panel.
        }                       %
\end{figure*}

We compare our map to similar results in the fields.
\citet{green20193d} reconstructed three quarters of the sky by combining Gaia and PANSTARRS data.
These data are combined in a Bayesian pipeline which also takes spatial correlations into account.
However, in their approach information is not as efficiently propagated along long distance, and therefore the map yields less constrained distance estimates.
The extinction estimates in general are expected to be better calibrated as in this map, as the likelihood was carefully constructed in their case.
Furthermore their sampling scheme is expected to yield superior uncertainty estimates.
A heatmap showing counts of pairs of integrated extinction for both maps is shown in \autoref{fig:heatmaps}.
Both maps are mostly well correlated, but less so for extinctions $A_G>0.5\,\text{mag}$.
We also show a comparison to our previous reconstruction in the bottom half of \autoref{fig:standard-deviation-comparison}.
We believe our previous reconstruction to represent the three dimensional structure of dust better, due to the reconstruction being less fragmented and due to the better statistical method employed for the reconstruction.
However, our new reconstruction has superior angular resolution and covers a far larger volume.
\autoref{fig:standard-deviation-comparison} also shows the standard deviation of our map as computed using the Fisher Laplace approximation.
We computed this standard deviation from the available two posterior samples provided by our algorithm and their antithetic counterparts, see \autoref{subsec:propagating-uncertainties} for more detail.

\section{Using the reconstruction}
\label{sec:using-the-reconstruction}

\subsection{Possible systematic errors}

When using the reconstruction one should be aware of certain systematic effects that should have degraded the result.
Namely we recommend only using the map between $-4\,\text{kpc}<x<2\,\text{kpc}$, $-4\,\text{kpc}<y<4\,\text{kpc}$, and $-1\,\text{kpc}<z<1\,\text{kpc}$ as there can be artifacts at larger distances.
We furthermore ask to be aware that clouds appear squashed in radial dimension on the map, leading to the appearance of circumsolar sheets of dust that in reality should be more plastic.
Furthermore there are artifacts at the edges of the cones, where two cones did not agree on the same distance for a dust cloud.

\subsection{Propagating uncertainties}
\label{subsec:propagating-uncertainties}

The reconstruction provides uncertainty quantification via two approximate posterior samples.
This is generally not enough to provide a good uncertainty estimate,
and the two sample also tend to underestimate the uncertainty, especially in the low-density regions.
We recommend to double the number of samples by mirroring the two samples around the maximum posterior estimate on log-scale, i.e. compute voxel by voxel
\begin{align}
    \rho_{i,\text{mirrored}} = \frac{\rho_\text{MAP}^2}{\rho_{i}} \ ,
\end{align}
where $\rho_i$ is the density of the $i$-th sample, and $\rho_\text{MAP}$ is the maximum posterior estimate.
This trick is known as antithetic sampling \citep{kroese2013handbook} and is known to reduce variance of estimators due to low sample number.
The reason why the mirroring is obtained by dividing $\rho_\text{MAP}$ by $\nicefrac{\rho_{i}}{\rho_\text{MAP}}$ is that the posterior statistics is Gaussian for logarithmic densities in our model.

Overall we recommend using the uncertainty estimates to distinguish artifacts from actual dust clouds as more of a sanity check and order of magnitude for uncertainty rather than as a proper uncertainty estimate.
The reconstruction has systematic errors that are not reflected by the statistical error budget as represented by the samples.

\subsection{Accessing the results}

The reconstruction has a total size of $1.4\,\text{TB}$.
In order to make the reconstruction better accessible, a portal to explore reconstruction interactively in three dimension and to obtain user specified parts of it has been deployed at \url{https://flute.esc.mpcdf.mpg.de/}, using software build for a cosmological web portal \citep{ragagnin2017web}. The site allows to query radial dust density slices as well as regularly spaced cuboids.

\section{Conclusion}
\label{sec:conclusion}

We present a large-scale map of the dust inside our corner of the Milky Way based on star reddening data and stellar parallaxes.
The reconstruction presented was based on using two new technological advances, namely a representation of a Gaussian process on an adaptive spherical grid and an improved handling of distance uncertainties.
Unfortunately, because of the large size of the reconstruction, only a maximum a posteriori approach could be afforded for the individually reconstructed cones and the information exchange between the cones could not be brought to full convergence.

Validation of the map using maser parallaxes and their association with reconstructed dust clouds indicates that the map is trustworthy up to $4\,$~kpc.
A number of reconstruction artifacts can be found, and hopefully these and others will be removed in future works.
These enclose sheet like dust filaments oriented perpendicular to the line of sights, which were caused by the used maximum a posteriori approach.
Furthermore, discontinuities at the boarders of the individual computational reconstruction cones can be observed despite our efforts to have the cones cross communicate.

Despite these shortcomings, we believe that the presented map is useful for many scientific purposes if handled with caution.
It will certainly be improved by future work, but until then, it might be a valuable representation of our current knowledge on the Galactic dust distribution in our Galactic quarter.


\begin{acknowledgements}

We acknowledge help by Klaus Dolag, Alexey Krukau, and Massin Guerdi for setting up our web portal.

This work has made use of data from the European Space Agency (ESA) mission
{\it Gaia} (\url{https://www.cosmos.esa.int/gaia}), processed by the {\it Gaia}
Data Processing and Analysis Consortium (DPAC,
\url{https://www.cosmos.esa.int/web/gaia/dpac/consortium}). Funding for the DPAC
has been provided by national institutions, in particular the institutions
participating in the {\it Gaia} Multilateral Agreement.

    This research was supported by the Excellence Cluster ORIGINS which is funded by the Deutsche Forschungsgemeinschaft (DFG, German Research Foundation) under Germany's Excellence Strategy – EXC-2094-390783311.

    We acknowledge the use of the Python packages NumPy \citep{harris2020array}, SciPy \citep{2020SciPy-NMeth}, Jax \citep{jax2018github}, hankel \citep{murray2019hankel}, Nifty5 \cite{arras2019nifty5}, Healpy \citep{Zonca2019,2005ApJ...622..759G}, and Matplotlib \citep{Hunter2007} to derive the results presented in this paper.

\end{acknowledgements}


\bibliography{ift}
\bibliographystyle{aa}

\appendix

\section{Grid Refinement for Angular Patches}
\label{sec:grid-refinement-for-cones}

\subsection{Coordinate definition}
\label{subsec:coordinate-definition}

We cover the sky with $424$ overlapping angular patches.
One patch covers an angular dimension of about $12.5^\circ\times12.5^\circ$.
All patches use their own local coordinate system which is rotated with respect to the Galactic coordinate system.
The location of the voxels with respect to the patches' own coordinate system is the same for each patch, only the rotation matrix that maps it to Galactic coordinates differs between patches.

For each patch, the coarsest coordinate system is indexed by $i\in\{0,\dots,19\}$, $j\in\{-3,\dots,3\}$, and $k\in\{-3,\dots,3\}$, leading to an initial grid of $19\times7\times7$ points.
We refine this grid $n$ times, with $n=6$ at the start of our optimization and $n=7$ at its end, using iterative grid refinement \citep{edenhofer2021sparse}.
Each refinement discards the outermost layer of voxels and splits each remaining voxel into 8 voxels.
This means that after $n=6$ refinements we have a grid of $964\times196\times196$ and at $n=7$ we reach our final resolution of $1924\times388\times388$.
A point $ijk$ after the $n$-th grid refinement corresponds to the location
\begin{align}
    r &= 40\cdot\text{exp}\left(\frac{\nicefrac{i-1.5}{2^n}}{2.5}\right) \label{eq:grid_def_r}\\
    l &= \phi_0 \nicefrac{j}{2^n}\\
    b &= \phi_0 \nicefrac{k}{2^n}\\
    \text{with } \phi_0 &= \frac{12.5^\circ}{7-3}\ ,
\end{align}
with $r$, $l$, and $b$ being radius, longitude, and latitude in the Sun-centered coordinate system of the corresponding patch.

\subsection{Refinement}
\label{subsec:Refinement}

We compute a separate refinement matrix $W$ for each distance bin and refinement level, each corresponding to a fixed $3\times3\times3$ convolution kernel.

This neglects distortions that are inherent to the spherical coordinate systems. To reduce these effects, we limited the cones to $12.5^\circ$, which leads to a maximal distortion of
\begin{align}
	1-\text{cos}\left(\frac{12.5^\circ}{2}\right) \approx 0.006\label{eq:distortion}\ ,
\end{align}
i.e. $0.6\%$. This is only a distortion in the prior correlation structure, and the likelihood is not affected by this. We believe the overall effect of this distortion to be negligible compared to other error sources.

We voxelize our reconstructed volume into equally spaced pixels in $l$ and $b$ while using logarithmic radial bins for $r$.
At larger distances at which the positional uncertainty increases, our radial resolution degrades.
Furthermore, as implied by \autoref{eq:grid_def_r} our grid starts at $\approx 40\ \text{pc}$, leaving out the innermost volume of our local bubble.
Thus, we effectively collapse the extinction of this inner sphere to the first layer of voxels in radial direction.
This effect is observed in our validation in \autoref{subsec:validation}.
For our main reconstruction we believe this effect to be negligible given our previous reconstruction \cite{leike2020resolving}) shows no dust in the innermost $40\,\text{pc}$.

The patches are arranged to cover the full sky.
At each equi-latitude ring $b=\{-90,-80,\dots 80,90\}$ there are $n_\text{patch}(b)$ patches, according to table\,\ref{table:patches-per-ring}.
A point $x$ in the coordinate system of the $u$-th patch at the $v$-th equi-latitude ring can be transformed to Galactic coordinates via the rotation matrix $M$ as
\begin{align}
    x^\prime &= Mx =
    \begin{pmatrix}
        \text{cos}(l) & \text{sin}(l) & 0\\
        \text{sin}(l) & \text{cos}(l) & 0\\
        0 & 0 & 1\\
    \end{pmatrix}
    \begin{pmatrix}
        \text{cos}(b) & 0 & \text{sin}(b)\\
        0 & 1 & 0\\
        \text{sin}(b) & 0 & \text{cos}(b)\\
    \end{pmatrix}x\\
    \text{with } b &= 10^\circ v\\
    \text{and } l &= \frac{360^\circ}{n_\text{patch}(b)}u,
\end{align}
where $x^\prime$ is the point $x$ in Galactic coordinates. In words, one first applies a rotation to bring a vector $(1,0,0)$ to latitude $b=10^\circ v$ and then rotates such that it points towards longitude $l=\frac{360^\circ}{n_\text{patch}(b)}u$.

\begin{table}
\centering
\footnotesize
\begin{tabular}{cc}
    latitude $b$ $[\text{deg}]$ &patches $n_\text{patch}(b)$\\
    \hline
    $\pm$90  &  1 \\
    $\pm$80  &  7 \\
    $\pm$70  &  13 \\
    $\pm$60  &  19 \\
    $\pm$50  &  24 \\
    $\pm$40  &  28 \\
    $\pm$30  &  32 \\
    $\pm$20  &  34 \\
    0,$\pm$10  &  36 \\
\end{tabular}
\normalsize
    \caption{Amount of patches per equi-latitude ring}
\label{table:patches-per-ring}
\end{table}

The refinement scheme described in \citep{edenhofer2021sparse} on the coordinate system defined in \autoref{subsec:coordinate-definition} enables us to enforce a physical correlation structure for the logarithm of the dust extinction density.
This scheme is an invaluable asset in the reconstruction of dust clouds using a physical prior.
The refinement iteratively adds small structures to a larger grid thus increasing the resolution of the grid.
It can best be understood for a $3\times3\times3$ grid of cubes, where the center cube is refined to $2\times2\times2$ grid of equal sized sub-cubes.
Given the Gaussian process realization $\tau(x_\text{coarse})$ at the centers $x_\text{coarse}$ of the cubes for the coarse $3\times3\times3$ grid, the Gaussian process realization $\tau(x_\text{fine})$ at the centers $x_\text{fine}$ of the cubes forming the $2\times2\times2$ fine grid are Gaussian distributed:
\begin{align}
    \tau(x_\text{fine}) \curvearrowleft \mathscr{G}(m_\text{fine}, D_\text{fine})\label{eq:grid-refinement-filter}\\
    \text{with } m_\text{fine} = W \tau(x_\text{coarse}) \\
\end{align}
Hereby $D$ and $W$ are uniquely determined by the Gaussian process kernel and the grid positions $x_\text{coarse}$ and $x_\text{fine}$.
Using the reparameterization trick we can write a sample of the Gaussian distribution in \autoref{eq:grid-refinement-filter} as
\begin{align}
    \tau(x_\text{fine}) = W\tau(x_\text{coarse}) + \sqrt{D}\xi^\prime,\label{eq:grid-refinement-generative}
\end{align}
where $\xi^\prime$ is a learned parameter and $\sqrt{D}$ is the matrix square root of $D$.
We apply this equation to all voxels of all grids at all refinement levels until we reach the desired refinement level, neglecting the dependence of $D$ and $W$ on $b$ and $l$.
Here we get the realization of the Gaussian process at refinement level $0$ by directly computing their covariance $D^{(0)}$ using the GP kernel, which due to the low amount of points at this level is analytically feasible.
This gives rise to a linear generative model for the logarithmic density $\tau$.

\subsection{Systematic errors}
\label{subsec:systematic-errors}

In the following, we will discuss multiple ways in which the refinement scheme introduces systematic errors in our Gaussian process prior.

There are two main sources for errors; one of which lies within the refinement step itself and the other is in the assumption underlying the refinement scheme.
Let us first start with the error in the refinement itself.
While refining the grid, $3 \times 3 \times 3$ coarse voxels are mapped to $2 \times 2 \times 2$ refined voxels centered around the central voxel of the coarse grid in a convolution like scheme.
By limiting the convolution kernel of the refinement to 3 voxels in each dimension, we drop information from points which are further away.
Furthermore by limiting ourselves to 2 voxels in each refined dimension, we neglect information from the added excitation to neighboring refined pixels.
These effects strictly increase the variance of the pixel itself and decrease correlations.

The second source of error lies in the iterative nature.
In each refinement it is assumed that the previous refinement level modeled the desired Gaussian process without error.
However, this assumption as outlined above does not hold.
The convolution kernel of the refinement mixes values of which the variance may be overestimated and the correlation underestimated.
From the first refinement level onwards, errors are introduced in the refinement and from the second level onwards these errors are mixed and potentially amplified.

The effect on the model of these error sources depends strongly on the kernel at hand.
Kernels which steeply fall off radially like the physical prior used here lead to smaller errors within a refinement compared to kernels with strong long-range correlations.

\begin{figure*}
        \begin{subfigure}[t]{.46\textwidth}
                \includegraphics[trim={0.7cm .2cm 1.5cm 1cm}, clip, width=.95\textwidth]{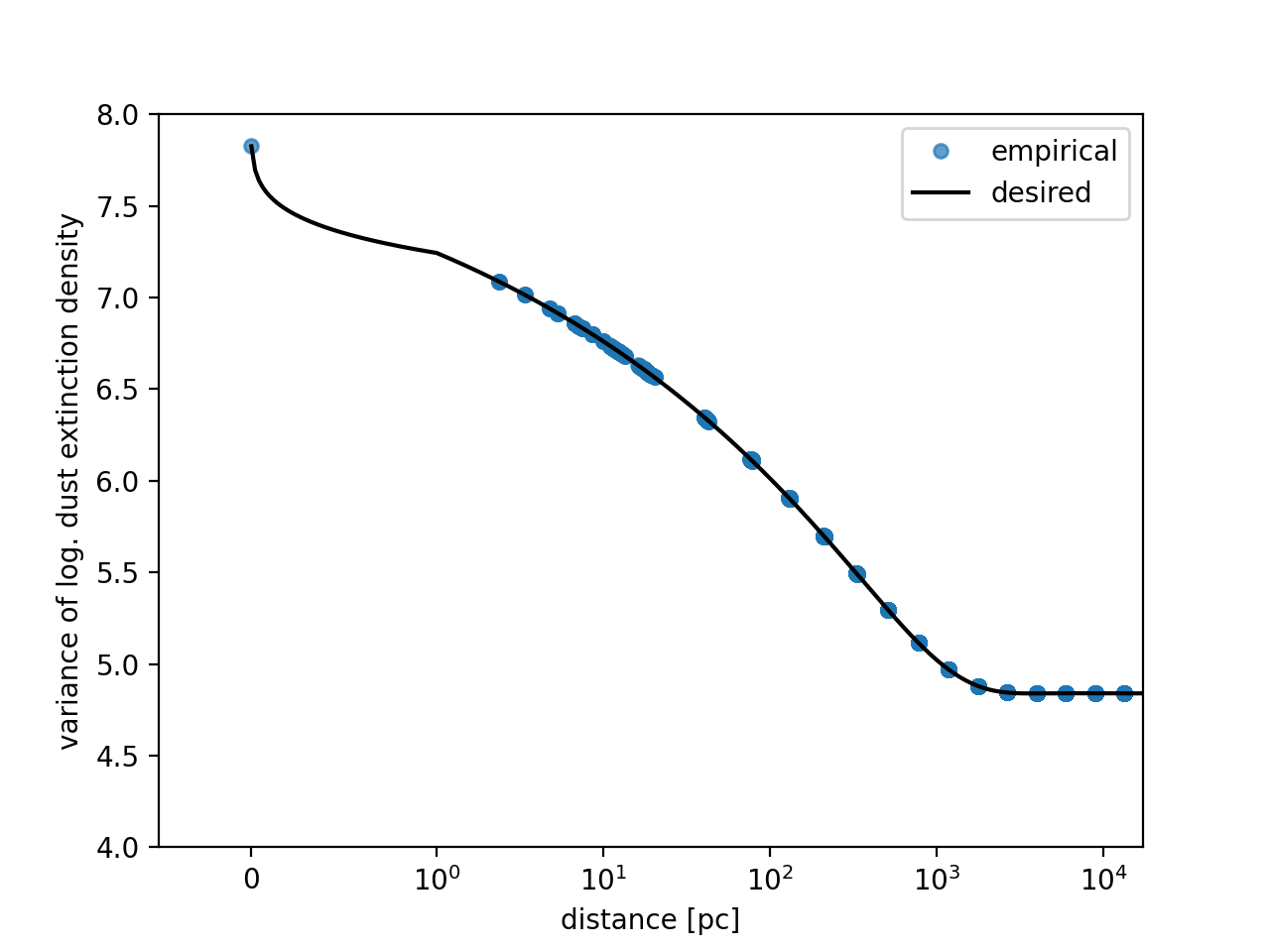}
        \caption{no refinement
        }
        \end{subfigure}~
        \begin{subfigure}[t]{.46\textwidth}
                \includegraphics[trim={0.7cm .2cm 1.5cm 1cm}, clip, width=.95\textwidth]{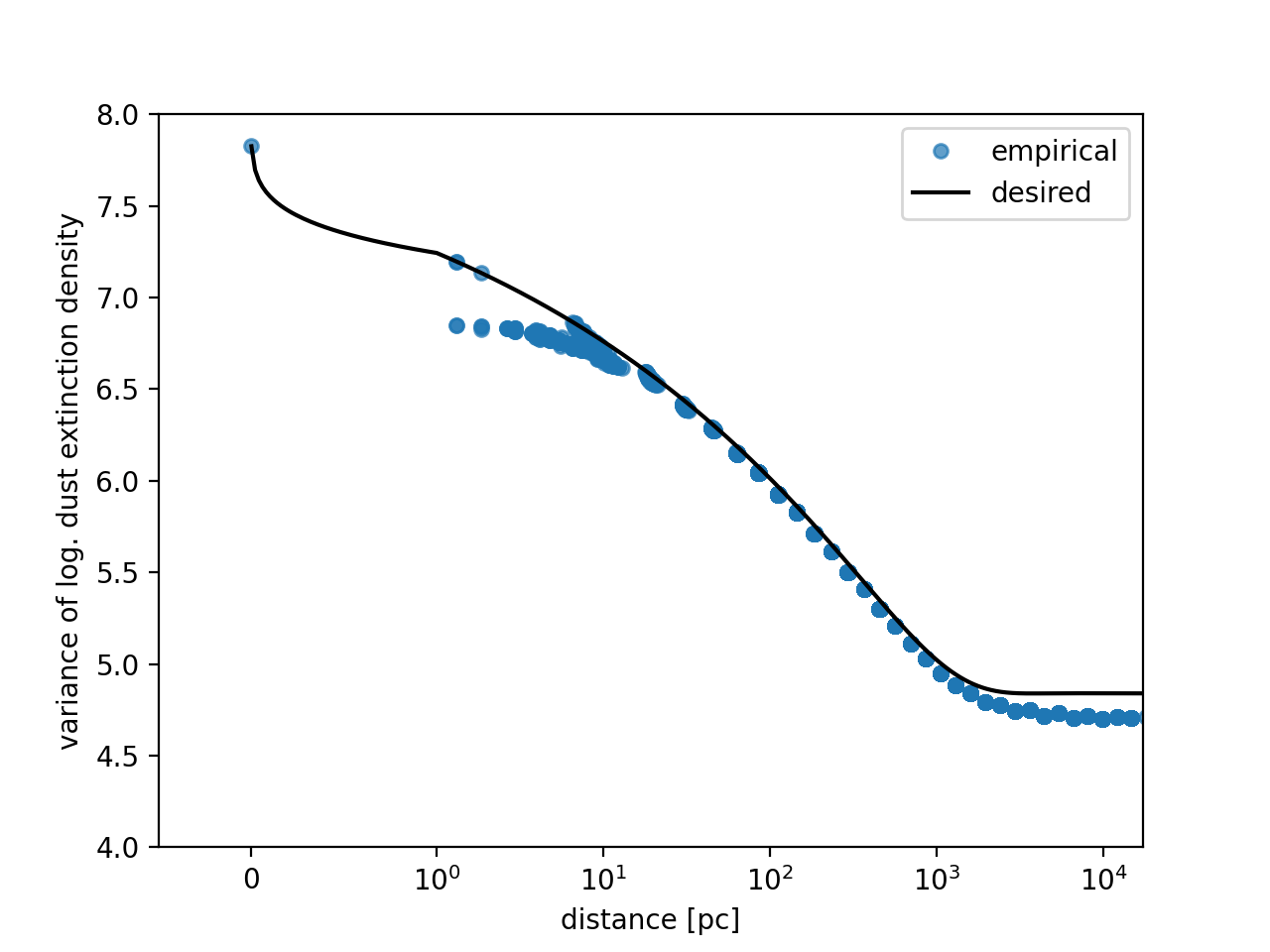}
        \caption{after first refinement
        }
        \end{subfigure}\\
        \begin{subfigure}[t]{.46\textwidth}
                \includegraphics[trim={0.7cm .2cm 1.5cm 1cm}, clip, width=.95\textwidth]{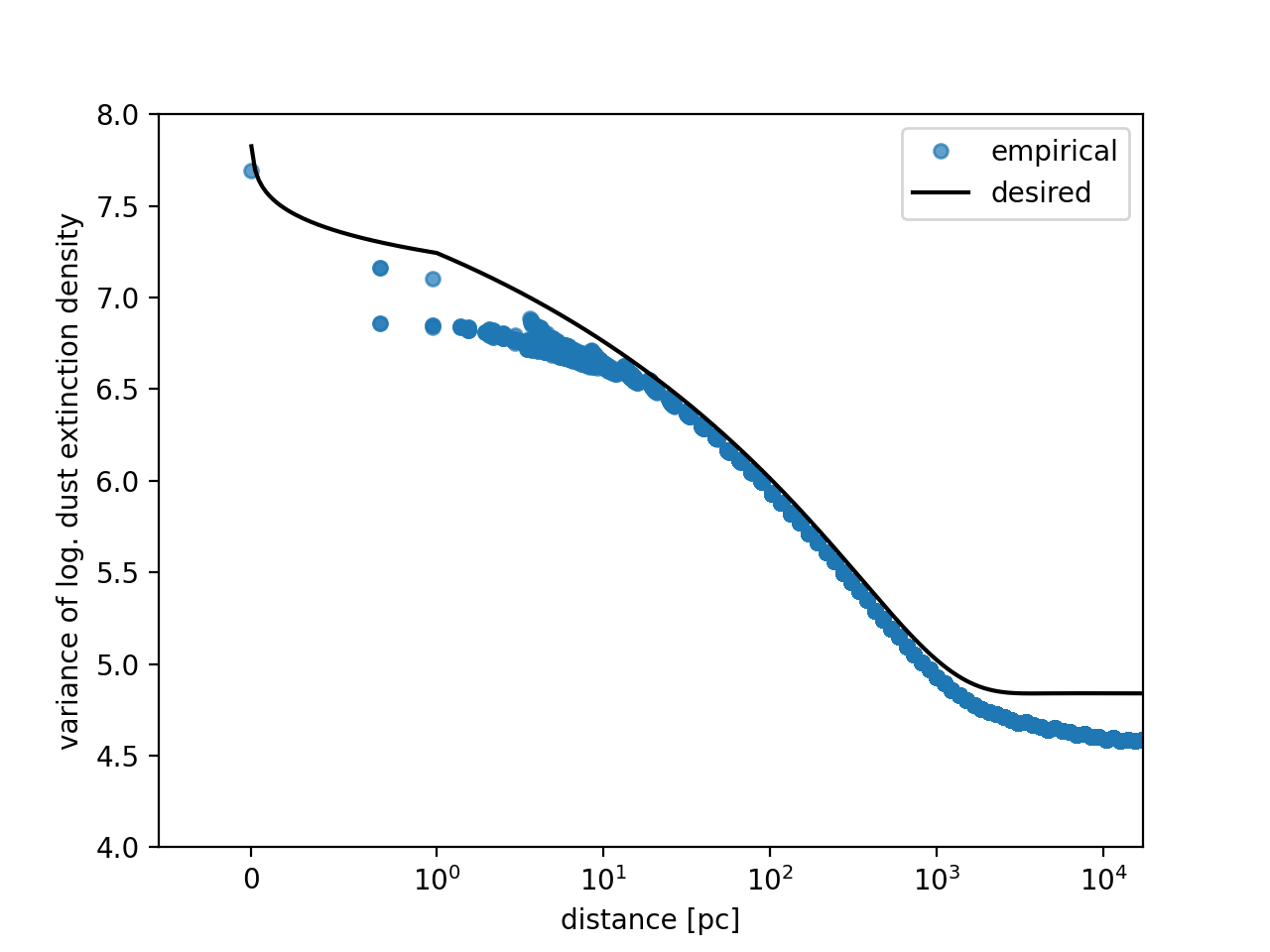}
        \caption{after second refinement
        }
        \end{subfigure}~
        \begin{subfigure}[t]{.46\textwidth}
                \includegraphics[trim={0.7cm .2cm 1.5cm 1cm}, clip, width=.95\textwidth]{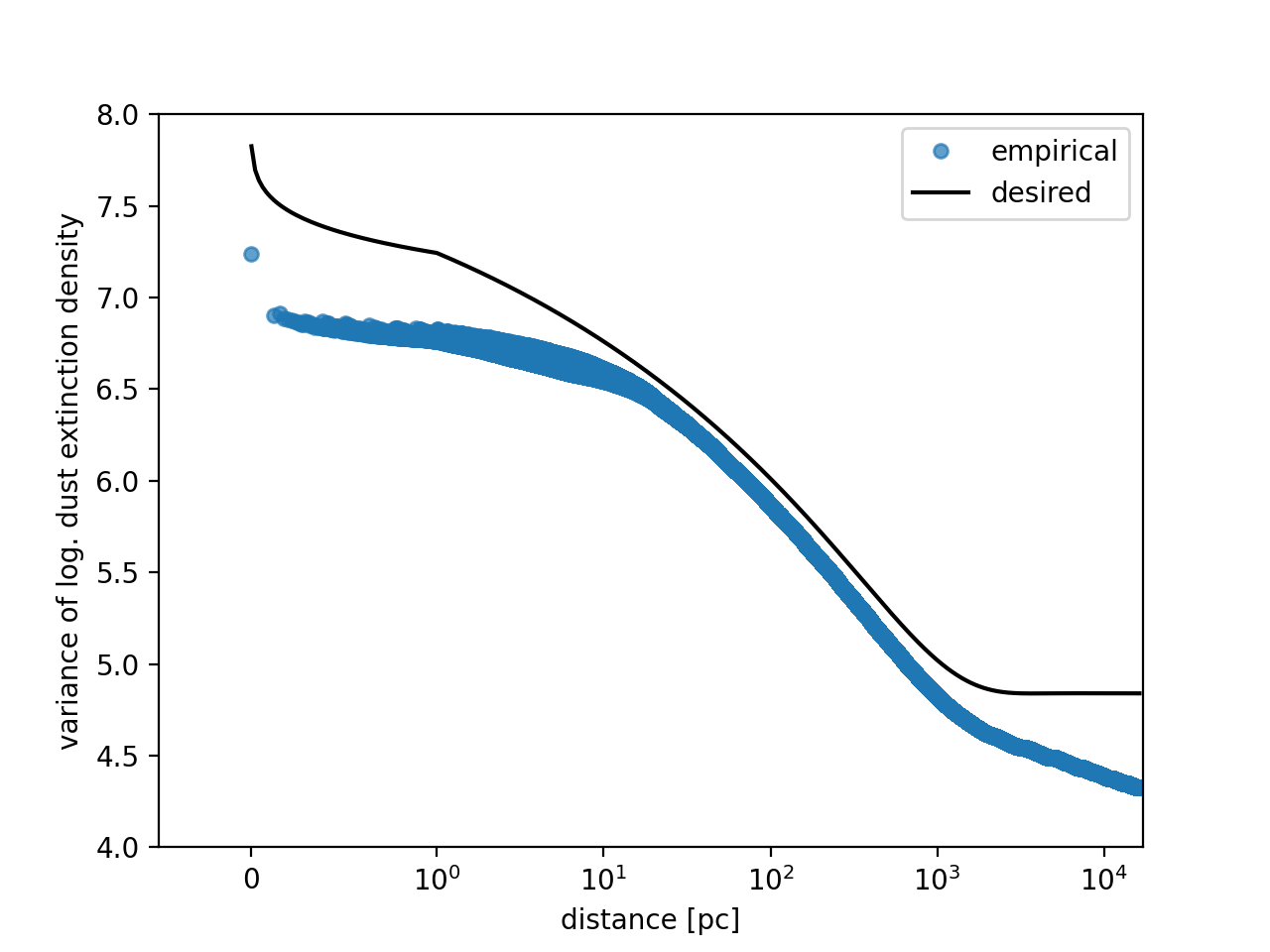}
        \caption{after final refinement
        }
        \end{subfigure}
        \caption{
        \label{fig:systematic-errors-in-log-dust-prior}
		The panels show the variance as a function of distance of the logarithm of the dust extinction density. The distance is taken relative to the first pixel at roughly $40\,\text{pc}$ from the Sun. The distance axis is linear in the interval $[0, 1]\,\text{pc}$ and logarithmic outside. The black line follows the desired prior correlation kernel. The blue dots depict the empirical kernel of the refinement scheme. Each dot represent an individual voxel in the grid at that refinement level.
        }%
\end{figure*}

\autoref{fig:systematic-errors-in-log-dust-prior} compares the empirical covariance kernel of the refinement scheme to the desired kernel.
At the zeroth refinement level the kernel is fit perfectly.
After the first refinement iteration the variance of the pixel is still correct as can be seen by the blue dot at distance zero.
However, correlations are dampened.
The second refinement further dampens correlations but now also the variance of voxels themselves is underestimated.
With increasing refinement level the errors accumulate and at the final refinement level the desired kernel is significantly below the desired one.

The magnitude of the errors additionally depends on the location in the grid.
This is because the refinement scheme depends on the distances between the to-be-refined voxels and the distances between voxels get larger the further away we are from the Sun.
In terms of Kullback–Leibler divergence, the loss of information comparing radial sub-slices of the true Gaussian process to our approximate one is the greatest at small distances to the Sun.

Overall our prior on the Guassian process using the refinement scheme systematically underestimates correlations and dampens the variance of voxels.
Even though we clearly underestimate the variance, being able to model Gaussian processes in linear time complexity for a grid of this size more than outweighs the here described systematic errors.
The refinement scheme models short- and long-range correlations with sufficient accuracy that it is a highly beneficial prior to enforce physical constraints for the logarithmic dust extinction density.

\section{Optimization and boundary conditions}
\label{sec:boundary-conditions}

We split the sky into 424 overlapping patches and start optimizing them independently of each other at half of our maximal resolution, i.e using 6 grid-refinements.
All patches are optimized until they are converged at that resolution level.
We define the result as converged if the logarithmic posterior probability density changes by less than $0.5$ for $3$ consecutive optimization steps or the parameters determining the highest resolution grid exceed $2$.
Once these thresholds are passed for all patches, after $1-7$ rounds of minimizing $12\,h$, we begin the final optimization at the target resolution using 7 grid-refinements, of which we performed $15$ rounds.

We assign each point in space the patch with the closest center, determined using $l$ and $b$ in local cone coordinates and an $\|.\|_\infty$ norm, i.e. we determine the cone in which the point has lowest $\text{max}(|l|,|b|)$.
Some points are covered by multiple patches, but all points are only assigned to a single patch, namely the closest one.
The closest patch is expected to host the most accurate reconstruction for that point.
The boundary of each patch is not assigned to the patch itself, but to a different patch.

We use this to propagate information between the cones.
Whenever a minimization job at the highest refinement level $n=7$ of a patch is started, we determine the patch assigned to each of the points $\tau_{ijk}$ on the radial boundary, i.e. with index $j=0$ or $k=0$ or $j=388$ or $k=388$.
We read the logarithm of the reconstructed differential extinction $\tau^\prime_{ijk}$ of the cone assigned to these points, respectively, and construct an additional likelihood
\begin{align}
	\mathscr{L} &= \prod_{i,j,k\,\in\,\text{boundary}}\mathcal{N}(\tau|\tau^\prime, \sigma)\\
    \text{with } \sigma &= 2^{6}
\end{align}
This additional loss term causes a penalty if the boundary of one patch does not agree with its neighbouring patches.

\begin{figure}
        \begin{subfigure}[t]{.46\textwidth}
                \includegraphics[trim={0.2cm .0cm 0.2cm .4cm}, clip, width=.95\textwidth]{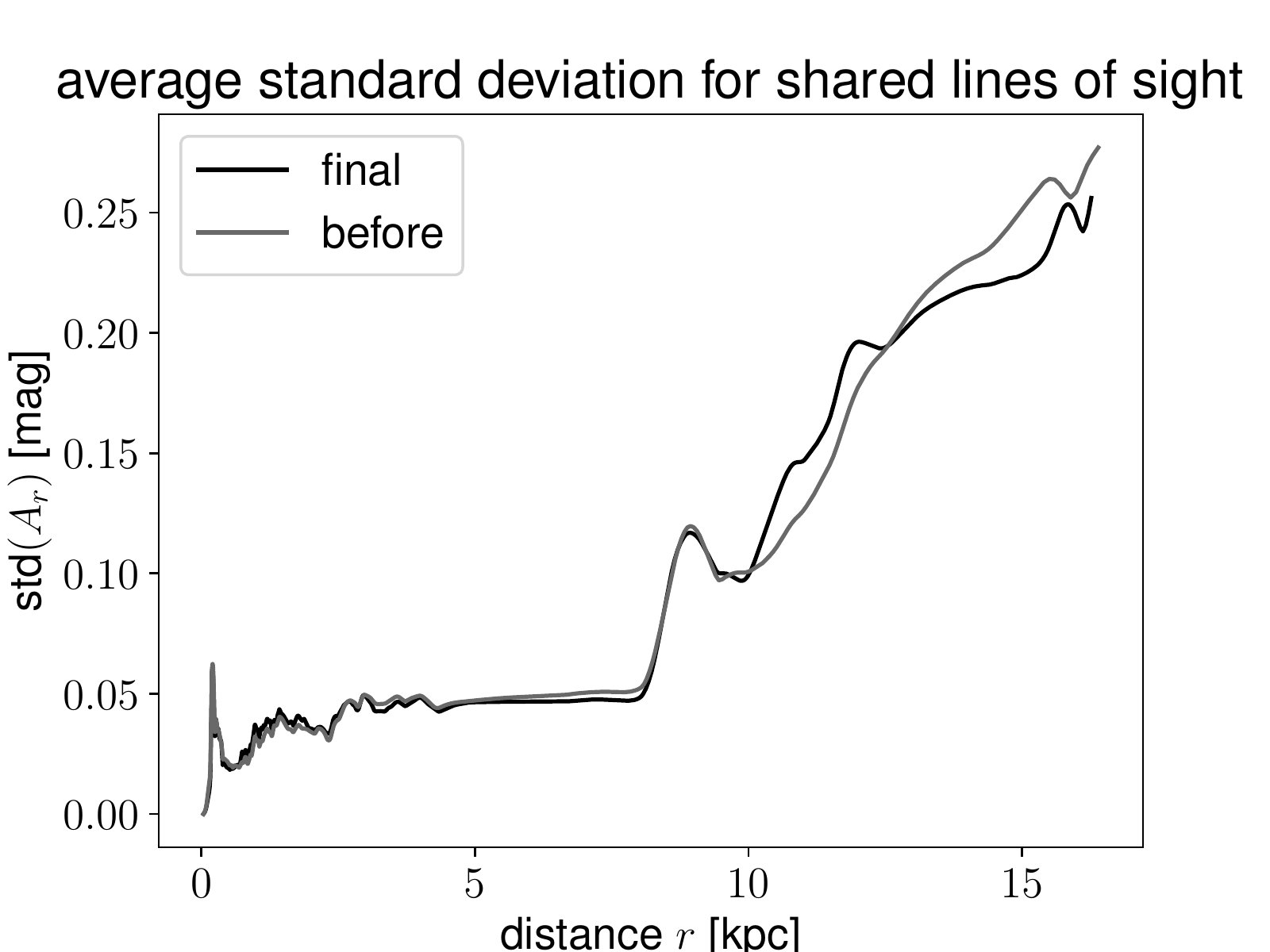}
        \caption{
        }
        \end{subfigure}
        \caption{
        \label{fig:overlap-mismatch}
        Mismatch of integrated extinction for lines of sight with longitude $l=5^\circ+10^\circ$ latitude $b=\pm4.981^\circ$, which are reconstructed by $4$ patches.
        }                       %
\end{figure}

Because the patches overlap for about $2.5^\circ$ but only the boundary needs to match, the final result can still disagree for large parts of the map.
We quantified this mismatch before starting the final minimization, where the patches are still independent, and after the final minimization.
The result can be found in Fig.\,\ref{fig:overlap-mismatch}.
Because the patches are minimized independently, they can converge to different local minima of our loss function given in \autoref{eq:overall-loss}, and thus the degree of their mismatch gives rise to an uncertainty estimate of the systematic error introduced by the maximum posterior approach.

\section{Fitting the kernel}
\label{sec:fitting-the-kernel}

To determine the parameters $(z_0,a,k_0,\alpha)$ of the kernel given by \eqref{eq:power-spectrum-form}
we use the Fourier transform of the logarithmic dust extinction samples of \citet{leike2020resolving}.
We compute bin averages of the logarithm of the Fourier transformed logarithmic dust extinction in logarithmically sized bins, ignoring the zero mode at $k=0$.
We then compute the least squares fit of the power law spectrum given by \eqref{eq:power-spectrum-form} to the bin averages, i.e. we minimize
\begin{align}
    \mathscr{L}_\text{P} &= \sum_i\left((\tau_i^2)^\text{avg}-\text{P}_k(a, k_0, \alpha)\right)\\
    \text{where } \tau_i^\text{avg} &= \text{avg}_{k_i<k<k_{i+1}}(\tau_k) \\
    \text{and } k_i &= 10^{-3}2^{\nicefrac{i}{2}}
\end{align}
We compute $z_0$ from the zero mode $P_0$ as
\begin{align}
    z_0 &= \frac{P_{0}}{V} \label{eq:define-z0}
\end{align}
where $V=(740\,\text{pc})^2(540\,\text{pc})$ is the total volume of the old reconstruction.
The correlation kernel $C(r)$ is then defined via the three dimensional inverse Fourier transform
\begin{align}
    C(r=\|x\|) = z_0 + \int\text{d}^3k\,\text{exp}(-ik^Tx)P_k \label{eq:define-correlation}
\end{align}

We treat the zero mode $z_0$, which corresponds to the expected a priori variance of the average logarithmic density, differently than the rest of the power spectrum.
As a reminder, the zero mode of the empirical power spectrum is defined as
\begin{align}
    P_0 = \frac{\tau_{k=0}^2}{V} &= \frac{1}{V}\left(\int \text{d}^3x\,\tau_x\right)^2 \label{eq:power-zero-mode}
\end{align}
i.e. it relates to the average logarithmic dust density per volume.
If we were to double the volume, then the $\frac{1}{V}$-term is there to cancel the volume dependence of $P_0$.
Canceling the volume dependence will work if the values $\tau_x$ takes in the additional volume $V^\prime$ are are independent to the values it takes in the original volume. 
We challenge this assumption, as we rather think that our models a-priori expected logarithmic dust extinction density is likely to have an offset to the true average dust density.
We believe this offset to be far larger than the intrinsic variations of the average dust density over the considered volume.
This offset is then expected to be the same in the additional subvolume, i.e. we will then get
\begin{align}
    \int_V \text{d}^3x_1\,\tau_{x_1} + \int_{V^\prime} \text{d}^3x_2\,\tau_{x_2} \approx
    2\int_V \text{d}^3x_1\,\tau_{x_1} .
\end{align}
This leads to a quadratic scaling of the zero mode with space, or a linear scaling of $P_0$ with space.
We want to make sure that the a-priori expected offset of the new reconstruction is the same as the average offset of the old reconstruction.
This is why we cancel this volume dependence in \autoref{eq:define-z0}.
Combining \autoref{eq:define-z0} and \autoref{eq:power-zero-mode} we get
\begin{align}
    z_0  &= \frac{1}{V^2}\left(\int \text{d}^3x\,\tau_x\right)^2\nonumber\\
    &= \left(\int \text{d}^3x\,\frac{\tau_x}{V}\right)^2\nonumber\\
    &= \overline{\tau}^2
\end{align}
where we defined $\overline{\tau}$ as the average logarithmic dust density.
From \autoref{eq:define-correlation} we get that
\begin{align}
    \lim_{r\rightarrow \infty}C(r) = z_0
\end{align}
meaning that the a-priori expectation of the product of two far away points $x_1$ and $x_2$ is $z_0$, exactly as is expected from two points that share a common expectation value $\overline{\tau}$ but are otherwise independent:
\begin{align}
    \mathbb{E}_{P(\tau_{x_1},\tau_{x_2})}\left(\tau_{x_1}\tau_{x_2}\right) &= \left(\mathbb{E}_{P(\tau_{x_1})}\tau_{x_1}\right)\left(\mathbb{E}_{P(\tau_{x_2})}\tau_{x_2}\right)\nonumber\\
    &= \overline{\tau}^2
\end{align}

Note that we can calculate $C(r)$ from $P_k$ using the Hankel transform instead of using \eqref{eq:kernel-from-power}, avoiding the need to perform a Fourier transform in three dimensions.
We use a Python implementation of the Hankel transform provided by \citet{murray2019hankel}.

\section{Alternative noise statistics}
\label{sec:alternative-noise}

Note that we use the G-band extinction $A_G$ of the \verb|StarHorse| data because we erroneous believed that this would reduce dependence of the extinction on the spectral type of the star.
However, the extinction $A_V$ is defined in the \verb|StarHorse| data set as extinction $A_\lambda$ at the wavelength $\lambda=542\,\text{nm}$, and not as extinction in the Johnson-Cousins V-band as the notation suggests.
Thus taking the $A_V$ data instead of $A_G$ could potentially slightly increase the quality of our reconstruction.
However, since we determine the likelihood given the extinction $A_G$, we do not expect to incur systematic errors from this decision.
To assess how much the reconstruction could have improved with the use of $A_V$ we fit a Student-T distribution to the 50-percentile of $A_V$ instead of $A_G$, but analogously to the procedure described above.
We find that the ratio of the scale parameter of the Student-T distributions found by the fit is consistent with the ratio $\nicefrac{A_V}{A_G}$, indicating no evidence that taking $A_V$ as data would have been more informative.

\end{document}